\documentclass[]{aa}

\usepackage[varg]{txfonts}
\usepackage{longtable}
\usepackage{bm}
\usepackage{graphicx}
\usepackage{color}
\usepackage{comment}
\usepackage{placeins}
\usepackage{capt-of}

\newcommand\kms{km~s$^{-1}$}
\newcommand\msun{$M_\odot$}
\newcommand\mhi{$M_{HI}$}

\def\be{\begin{equation}}
\def\ee{\end{equation}}
\def\a40{$\alpha$.40}

\def\arcsec{$^{\prime\prime}$}
\def\dg{$^{\circ}$}
\def\w50{$W_{50}$}

\def\hi{H\,{\sc i}}
\def\mstar{$M_{*}$}
\def\sbunit{mag arcsec$^{-2}$}

\def\bm{beam$^{-1}$}
\def\mueffg{$\overline{\mu}_{\mathrm{eff,g}}$}
\def\muo{$\mu_{0,g}$}
\def\re{$r_{\mathrm{eff}}$}

\usepackage{natbib,twoopt}
\usepackage[breaklinks=true,draft]{hyperref} 
\bibpunct{(}{)}{;}{a}{}{,} 
\makeatletter
\newcommandtwoopt{\citeads}[3][][]{\href{http://adsabs.harvard.edu/abs/#3}%
{\def\hyper@linkstart##1##2{}%
\let\hyper@linkend\@empty\citealp[#1][#2]{#3}}}
\newcommandtwoopt{\citepads}[3][][]{\href{http://adsabs.harvard.edu/abs/#3}%
{\def\hyper@linkstart##1##2{}%
\let\hyper@linkend\@empty\citep[#1][#2]{#3}}}
\newcommandtwoopt{\citetads}[3][][]{\href{http://adsabs.harvard.edu/abs/#3}%
{\def\hyper@linkstart##1##2{}%
\let\hyper@linkend\@empty\citet[#1][#2]{#3}}}
\newcommandtwoopt{\citeyearads}[3][][]%
{\href{http://adsabs.harvard.edu/abs/#3}
{\def\hyper@linkstart##1##2{}%
\let\hyper@linkend\@empty\citeyear[#1][#2]{#3}}}
\makeatother

\begin{document}

\title{MIGHTEE \hi\ observations of low surface brightness and ultra-diffuse galaxies in the XMM-LSS field}

\author{Elizabeth A.\ K.\ Adams 
\inst{1,2}
\and 
Barbara Šiljeg
\inst{1,2}
\and
Anastasia A.\ Ponomareva
\inst{3,4}
\and
Natasha Maddox
\inst{5}
\and
Pavel E. Mancera Pi\~na
\inst{6}
\and 
Marten Baes
\inst{7}
\and
Bradley Frank
\inst{8}
\and
Marcin Glowacki
\inst{9}
\and
Matt J. Jarvis
\inst{10,11}
\and
Sambatriniaina H.\ A.\ Rajohnson
\inst{12}
\and 
Gauri Sharma
\inst{13,14,15,16,11}
}
\institute{
ASTRON, Netherlands Institute for Radio Astronomy, Oude Hoogeveensedijk 4, 7991 PD Dwingeloo, The Netherlands
 \email{adams@astron.nl}
\and
		Kapteyn Astronomical Institute, University of Groningen, Postbus 800, 9700 AV Groningen, The Netherlands
\and
Centre for Astrophysics Research, School of Physics, Astronomy and Mathematics, University of Hertfordshire,
College Lane
\and
Astrophysics, Department of Physics, University of Oxford, Keble Road, Oxford OX1 3RH, UK
\and
School of Physics, H.H. Wills Physics Laboratory, Tyndall Avenue, University of Bristol, Bristol, BS8 1TL, UK
\and
Leiden Observatory, Leiden University, P.O. Box 9513, 2300 RA, Leiden, The Netherlands
\and
Sterrenkundig Observatorium, Krijgslaan 281 S9, B-9000 Gent, Belgium
\and
STFC UK Astronomy Technology Centre, Royal Observatory, Edinburgh, Blackford Hill, Edinburgh, EH9 3HJ
\and
Institute for Astronomy, University of Edinburgh, Royal Observatory, Edinburgh EH9 3HJ, UK
\and
Astrophysics, Department of Physics, University of Oxford, Keble Road, Oxford, OX1 3RH, UK
\and 
Department of Physics and Astronomy, University of the Western Cape, Robert Sobukwe Road, 7535 Bellville, Cape Town, South Africa
\and 
INAF - Osservatorio Astronomico di Cagliari, Via della Scienza 5, I-09047 Selargius (CA), Italy
\and
University of Strasbourg, CNRS UMR 7550, Observatoire astronomique de Strasbourg, F-67000 Strasbourg, France
\and  
SISSA International School for Advanced Studies, Via Bonomea 265, I-34136 Trieste, Italy
\and  
QGSKY, INFN-Sezione di Trieste, via Valerio 2, I-34127 Trieste, Italy
\and  
IFPU Institute for Fundamental Physics of the Universe, Via Beirut, 2, 34151 Trieste, Italy
}

\titlerunning{MIGHTEE LSB detections}

\abstract{
Untargeted neutral hydrogen (\hi) surveys are well suited to identifying low surface brightness galaxies (LSBGs) that are gas rich, and they offer a complementary view to optically selected populations. We examined the LSBG population as identified via stellar and gaseous content using the MIGHTEE \hi\ XMM-LSS early science data and the publicly available catalogs of optically identified LSBGs.
There is currently little overlap between these datasets, with only three galaxies commonly detected. We performed surface brightness photometry of selected MIGHTEE \hi\ detections to find 29 LSBGs, and 26 of these meet the size requirement ($R_{eff} > 1.5$ kpc) to be ultra-diffuse galaxies (UDGs).
Furthermore, we extracted \hi\ spectra at the location of all optically identified galaxies, placing upper limits on the \hi-to-stellar mass ratio in these systems.  
While the \hi-identified population overall tends toward bluer colors, the \hi-identified and the optically selected samples mostly overlap in mean effective surface brightness, effective radii, and color. Although it is not straightforward to discern why the \hi-identified LSBGs were missed in optical searches, this work highlights the utility of \hi\ surveys in finding these faint systems. The \hi-identified LSBGs are gas rich compared to the general \hi-selected population. Furthermore, three out of four \hi-selected UDGs with available kinematics show no systematic offset from the baryonic Tully-Fisher relation, although we are biased away from sources with low rotational velocities due to the low spectral resolution of the data. This work demonstrates the utility of \hi\ observations for finding and characterizing the low surface brightness Universe.
}


\maketitle


\section{Introduction}\label{sec:intro}

Low surface brightness galaxies (LSBGs)
have
been long known and recognized for their importance in constraining galaxy formation models \citepads[e.g.,][]{1984AJ.....89..919S, 1997ApJ...482..659D, 1988ApJ...330..634I}. Interest in the low surface brightness (LSB) universe was reinvigorated with the discovery of thousands of very extended LSBGs in the Coma Cluster, which are referred to as ultra-diffuse galaxies (UDGs) and were characterized by their LSB nature and 
large effective radii \citepads[][]{2015ApJ...798L..45V, 2015ApJ...807L...2K, 2016ApJS..225...11Y}. The realization, enabled by optical surveys with increased sensitivity, that there was a large population of previously overlooked galaxies spurred observers to search for UDGs and theorists to explain their formation mechanism.

The original identification of UDGs in clusters and their need to survive that environment motivated formation theories where UDGs were "failed" galaxies that lived in massive dark matter halos and were quenched by early infall into the cluster \citepads[e.g.,][]{2015MNRAS.452..937Y}. 
Other work identified the cluster UDGs as a subset of the cluster dwarf galaxy population, with their larger sizes attributable to interactions within the cluster \citepads[e.g.,][]{2017A&A...608A.142V,2018RNAAS...2...43C, 2019MNRAS.485.1036M, 2023MNRAS.521.2012W}. At the same time, other models have predicted that UDGs are a natural part of the dwarf galaxy population in the field,
for instance, that they are the tail of the high-spin halo population \citepads[e.g.,][]{2016MNRAS.459L..51A,2017MNRAS.470.4231R,2018MNRAS.475..232P,2020MNRAS.495.3636M}, they form due to gas outflows from star formation \citepads{2017MNRAS.466L...1D}, or they result
from early major mergers or collisions \citepads{2021MNRAS.502.5370W,2019MNRAS.488L..24S}. Testing these theories requires identifying UDGs in the field population and connecting them to the cluster population.

When searching for field UDGs  in optical surveys, redshifts are generally not available, as by their nature these sources are too faint
for most large-scale redshift surveys. Thus, selection criteria are based on surface brightness (distance-independent) and angular size.
For example, \citetads[][hereafter G18]{2018ApJ...857..104G} searched the wide layer of the Hyper Suprime-Cam (HSC) Subaru 
Strategic Program (SSP) for galaxies with an effective radius of \re$= 2.5-14$\arcsec and an average surface brightness within \re\ in the $g$ band \mueffg $>24.3$ mag arcsec$^{-2}$. Similarly,
the team of Systematically Measuring Ultra-diffuse Galaxies (SMuDGEs) looked at the Legacy Surveys to find galaxies with a peak surface brightness of \muo $\ge  24$ mag arcsec$^{-2}$ and \re$\ge 5.3$\arcsec\citepads{2023ApJS..267...27Z}.
These provided LSBGs, some of which were UDG candidates. 
However, the UDG classification relies on physical size, in particular $R_{\text{eff}}$ > 1.5 kpc, and thus it requires a distance estimate.
Approaches to estimating distances have included statistically correlating with known overdensities or galaxies with known redshifts \citepads{2022ApJS..261...11Z,2022ApJ...933..150G}, focusing on satellite galaxies associated
with a host with a known redshift \citepads[][hereafter L23]{2023ApJ...955....1L}, or
using either the surface brightness fluctuations \citepads{2019ApJ...879...13C} or the globular cluster luminosity function \citepads[e.g.,][]{2019MNRAS.486..823R}.

\citetads{2017ApJ...842..133L} and \citetads{2019MNRAS.490..566J}
showed that there is a large population of gas-rich UDGs also present in neutral hydrogen (\hi) surveys. In fact, these studies found gas-bearing UDGs to be extremely gas rich for their stellar mass, with gas fractions (\mhi/\mstar) of up to 100.
An \hi\ detection is completely 
independent of the stellar content, enabling detection of galaxies that might be missed in optical searches due to insufficient sensitivity or biases in detection methods (e.g., due to shredding). Hence, it provides a complementary view to optical searches for LSBGs and UDGs.

In addition to untargeted \hi\ searches, 
targeting optically selected LSBGs in \hi\ provides a means to obtain redshifts and test their UDG classification,
as was done for SMuDGEs \citepads{2024ApJ...975...91K}. With the wealth of new \hi\ surveys and facilities available in preparation for the SKA-Mid telescope, the opportunities to combine optical and \hi\ data for understanding LSBGs and UDGs are expanding. For example, \citetads{
2023MNRAS.526.3130F} searched pre-pilot data from the Widefield ASKAP L-band Legacy All-sky Blind surveY (WALLABY) for detections of SMUDGES galaxies, finding one \hi-rich LSBG and non-detections of \hi\ in six putative UDGs.

In this work, we
combine the optical and \hi\ views of LSBGs and UDGs. 
The Early Science (ES) \hi\ data from the MeerKAT International Giga Hertz Tiered Extragalactic Exploration {MIGHTEE} survey \citepads{2016mks..confE...6J} provide a catalog of \hi\ detections, and previous optical searches provide catalogs of LSBGs and UDG candidates; we introduce these catalogs in Section \ref{sec:data}.
We discuss how we cross-matched the catalogs in Section \ref{sec:crossmatch} and detail the new LSBGs and UDGs we identified from the \hi\ detections in Section \ref{sec:hi_lsb}. In Section \ref{sec:hi_content} we explore the \hi\ content of optically identified galaxies. We discuss the implications of our results in Section \ref{sec:discussion} and end with our conclusions in Section \ref{sec:conclusions}.

\begin{figure*}[t]
    \sidecaption
    \includegraphics[width=12cm,keepaspectratio]{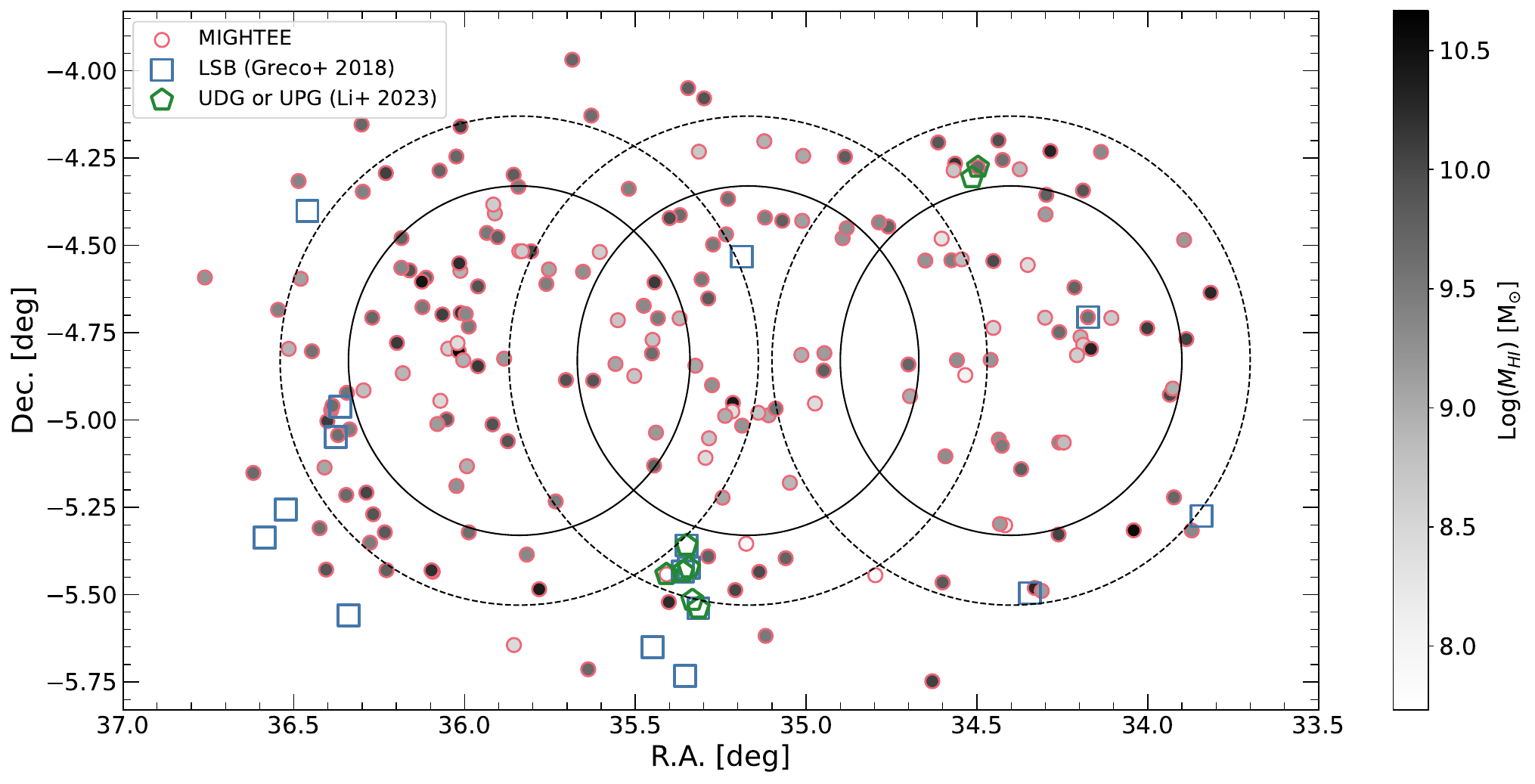}
    \caption{Region of XMM-LSS covered in the MIGHTEE ES data. From left-to-right (decreasing right ascension), the three fields are XMM-LSS-14, XMM-LSS-13, and XMM-LSS-12. The solid lines indicate the full width half maximum of the MeerKAT primary beam, and the dashed lines represent the 30\% response level. The MIGHTEE \hi\ detections are shown as circles, shaded based on their \hi\ mass. Also included are the LSBGs of G18 (blue squares) and the UDGs and UPGs of L23 (green pentagons) that fall within the considered footprint.
    }
    \label{fig:xmm}
\end{figure*}

\section{Data}\label{sec:data}
The field targeted by XMM-Large Scale Structure (XMM-LSS) survey is a well-studied extragalactic field that
is contained within both HSC-SSP and MIGHTEE. For the purpose of this work, we focus on a subset of the region that contains the MIGHTEE ES footprint,  
shown in Figure \ref{fig:xmm}.

\subsection{MIGHTEE}\label{sec:mightee}

\subsubsection{\hi\ data}
The MIGHTEE ES data 
include
three pointings in the XMM-LSS field. These pointings are referred to as XMM-LSS-12, XMM-LSS-13, and XMM-LSS-14 and are shown in Figure \ref{fig:xmm}.
The MIGHTEE ES data covers 1310-1420 MHz (z=$0-0.084$) with
a spectral resolution of 208.984 kHz ($\sim$44 \kms at z=0).
For logistical reasons, the bandwidth is split into three congruent spectral windows for processing.
The typical noise is 81 $\mu$Jy \bm\ in a single channel.
Within a single channel, this
corresponds to a 3$\sigma$ column density sensitivity
of 9.8 $\times 10^{19}$ atoms cm$^{-2}$. 

The MIGHTEE ES \hi\ source catalog was constructed via unguided visual inspection of the cubes to identify \hi\ emission \citepads{2021A&A...646A..35M}. The \hi\ masses were derived from measurements of the integrated flux density in moment-0 maps \citepads{2021MNRAS.508.1195P, 2022MNRAS.512.2697R}. The \hi\ velocity widths at 50\% of the peak flux density (\w50) were derived by fitting the busy function \citepads{2014MNRAS.438.1176W,2021MNRAS.508.1195P}. Distances for the MIGHTEE ES \hi\ catalog are cosmological, with an assumed cosmology of
H$_{0}$ = 67.4 km s$^{-1}$ Mpc$^{-1}$ , $\Omega_{m}$ = 0.315,
and $\Omega_{\Lambda}$= 0.685 \citepads{2020A&A...641A...6P}.
However, we adopt the distances based on local flow models for our analysis of LSBGs and UDGs, as described in Section \ref{sec:surf_bright_phot}. 
The 196 detections in the XMM-LSS ES region are shown
in Figure \ref{fig:xmm}.

\subsubsection{Ancillary MIGHTEE data}\label{sec:ancillary}
The MIGHTEE fields were chosen specifically for their wealth of ancillary data, including deep optical photometry from HSC and deep near-infrared photometry from VISTA, as detailed in \citetads{2021A&A...646A..35M}.
In particular, the XMM-LSS field has either deep or ultra-deep HSC-SSP data in $u, g, r, i,$ and $z$, which is $\sim$0.5--1.5 mags deeper than the wide tier with the 5$\sigma$ point-source detection
limits for the deep (ultra-deep) data of $g = 27.5$ (28.1), $r=  27.1$ (27.7), and $i = 26.8$ (27.4) magnitudes \citepads{2018PASJ...70S...4A}.

The ancillary data were used to provide value-added information about the 
MIGHTEE \hi\ detections. Elliptical apertures whose purpose was to encompass as much
of the light as possible in the HSC $g$ band image were applied 
to the photometric data from the $u$ band to the NIR and used in the spectral energy distribution (SED) fitting to derive stellar masses and star formation rates \citepads{2021A&A...646A..35M}.
Thus, we also have the size of the elliptical aperture used for photometry, the $g$ and $i$ band magnitudes, and the stellar masses for all MIGHTEE ES \hi\ detections.

\begin{table}
\centering
\caption{\hi\ and optical cross-matched sources.}
\begin{tabular}{lllll}
\hline \hline
Name & Optical ID\tablefootmark{a} & Sep. & z$_{HI}$ & z$_{host}$ \\
MGTH\_ &  & $^{\prime\prime}$ &  &  \\
\hline
J021642.0-044221 & G267 & 0.6 & 0.042752 &  \\
J021759.3-041634 & L25657 & 0.7 & 0.033398 & 0.035 \\
J022138.1-052631 & L86749 & 1.5 & 0.006912 & 0.032 \\
\hline
\end{tabular}
\tablefoottext{a}{"G" refers to G18 and "L" to L23}
\label{tab:xmatch}
\end{table}

\begin{figure*}
    \centering
    \includegraphics[width=0.9\linewidth]{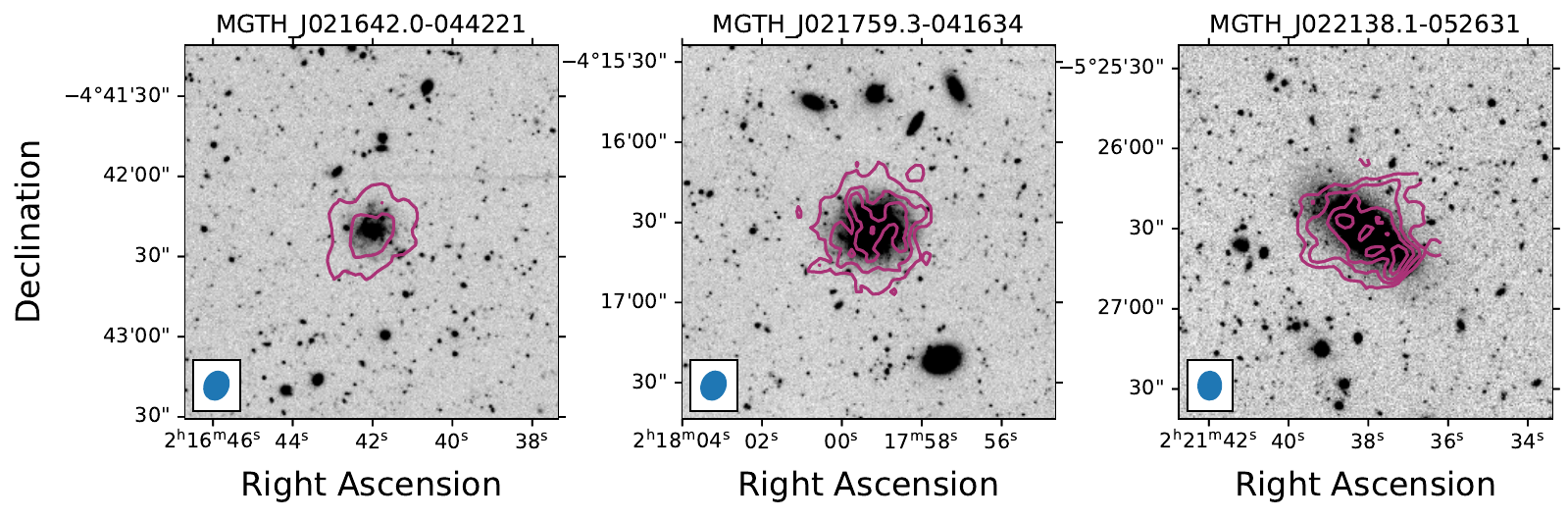}
    \caption{The three cross-matched optical and \hi\ sources. \hi\ contours at [2.5, 5, 7, 10] $\times 10^{20}$ atoms cm$^{-2}$ overlaid on $g$ band images. 
    }
    \label{fig:overlay_matches}
\end{figure*}

\subsection{Optically selected LSBGs}

We used the catalogs of G18 and L23 to select optically identified LSBGs within the MIGHTEE ES XMM-LSS footprint.
Both works use the wide layer of the HSC-SSP,
which covers 
five broad
bands ($g, r, i, z, y$) with a depth of
$g = 26.6$ mag, $r = 26.2$ mag, and $i = 26.2$ mag (5$\sigma$ point-source detection) and a median seeing of 0.6\arcsec\ \citepads{2018PASJ...70S...4A}.
The main difference between the two catalogs is that
G18 used the internal data release S16A, while
L23 used the public data release 2 \citepads[PDR2; also
known as S18A][]{2019PASJ...71..114A}, which provides sky subtraction
that is more appropriate for LSBGs.

Both works identified LSBGs, albeit in two different ways.
In G18,
masked galaxy cutouts were modeled
as a two-dimensional point spread function (PSF)-convolved Sérsic function.
Sources with \mueffg$> 24.3$ and 2.5\arcsec\ < \re\ < 14\arcsec\ were identified as LSBGs, and they could be UDGs 
depending upon their (unknown) distance.
In L23, a Spergel surface brightness profile  \citepads{2010ApJS..191...58S,2024A&A...684A..23T} was convolved with the PSF to model the galaxies in the catalog. L23 also used a deblending technique that improved their completeness. The authors of L23 matched 
LSBGs to satellites of Milky Way-mass host galaxies
from the NASA-Sloan Atlas,\footnote{\url{https://nsatlas.org/}} which provides compiled spectroscopic redshifts or direct distance measurements. They assigned the host redshift to the LSBG
and then only
provide catalogs for the galaxies that are either
UDGs ($R_{eff} + \sigma(R_{eff}) > $ 1.5 kpc and \mueffg + $\sigma$(\mueffg) $> 25.0$ \sbunit\, 
in their definition) or ``ultra-puffy galaxies'' (UPGs), galaxies that are outliers by more than 1.5$\sigma$ from the mass-size relation defined by satellites
of Milky Way analogs in the Local Volume \citepads{2021ApJ...922..267C}.
Thus, the L23 catalog has higher quality data with additional information (host redshift) but only for a subset of satellite galaxies compared to the G18 catalog. Given the different strengths of these two catalogs, we utilized both. Within the region of sky we considered (see Figure \ref{fig:xmm}), there are 16 LSBGs from G18. The catalogs of L23 provide eight galaxies, five of which are classified as both UDG and UPG, two as UDGs only, and one as UPG only. Of the eight galaxies in L23, four are newly cataloged compared to G18. Given the improvements of L23 and additional information, we used measured properties from L23 over G18 when sources are cataloged in both. The galaxies from these catalogs are shown in Figure \ref{fig:xmm}.

\section{Cross-matching \hi\ and optical}\label{sec:crossmatch}
As a starting point, we cross-matched
the existing MIGHTEE ES \hi\ source list
with the optically identified LSBGs of
G18 and L23
with a 5\arcsec\ tolerance,
finding three sources in common.  
These three sources are shown in Figure \ref{fig:overlay_matches}, where the MIGHTEE \hi\ contours are plotted on (ultra-)deep $g$ band HSC data. 
We list these sources in Table \ref{tab:xmatch} along with their angular separation and both the MIGHTEE \hi\ and host redshifts (if available).

The source MGTH\_J021642.0-044221 is in G18 and thus had no existing redshift information available before.
The source MGTH\_J021759.3-041634 is in the UPG catalog of L23 but not in the UDG catalog.
The \hi\ redshift is consistent with the assigned host galaxy redshift of 0.035.
The source MGTH\_J022138.1-052631 is in both the UPG and UDG catalogs of L23. However, the \hi\ redshift (0.0069) is significantly lower than that of the assigned host (0.032),
implying an incorrect host assignment.
Thus this galaxy is much closer and has a much smaller
physical effective radius than determined by L23. Based on the MIGHTEE redshift and cosmological distance, we find the physical effective radius (based on the L23 measured effective radius) is only 0.797 kpc, as opposed to the 3.594 kpc found by L23. Thus it would no longer be considered a UDG nor a UPG (for the updated stellar mass of log(\mstar/\msun) = 6.7).

\section{\hi-identified LSB galaxies}\label{sec:hi_lsb}

While there were not many direct cross-matches between the MIGHTEE \hi\ detections and the optically identified LSBGs, \hi\ is often an excellent tracer for finding LSBGs overlooked in the optical. Thus, we took the complementary approach of searching for LSBGs within the MIGHTEE \hi\ detections.

\subsection{Selecting potential LSBGs}

To identify LSBGs, we performed surface
brightness photometry of a subset of MIGHTEE \hi\ detections. With 196
total detections in the XMM-LSS MIGHTEE ES field, we preselected the
most promising LSB targets for surface brightness photometry following two different approaches.

The MIGHTEE ES XMM-LSS \hi\ source list includes aperture photometry for all
detections within the footprint of the deep or ultra-deep HSC data
at the time of its creation \citepads{2021A&A...646A..35M}. There are five
\hi\ sources that fell outside the deep HSC coverage at this time, and they are not considered further in this work.
In order to select targets for surface brightness photometry, we approximated the effective radius as one third of the total size of the aperture used for photometry\footnote{Elliptical apertures were individually set to contain all the light of each galaxy for the global photometry of the MIGHTEE \hi\ ES catalog, as described in Section \ref{sec:ancillary}. 
{For a typical S\'{e}rsic index of $n=1$, the ratio of \re\ to the outermost radius (containing $95-99$\%) of the light is approximately three \citepads{2005PASA...22..118G}}.} and used this to estimate \mueffg. 
There were 19 sources with
an estimated \mueffg$ > 24.3$, and we performed surface brightness photometry of these targets 
to accurately measure their \re\ and \mueffg\ in Section \ref{sec:surf_bright_phot}. 

In addition, we undertook visual inspection of 
the $g$ band fits images for all MIGHTEE XMM-LSS \hi\ detections with (ultra-)deep HSC data to identify potential LSBGs for surface brightness photometry. This added an additional 14 targets, in addition to re-identifying seven of the automatically selected targets. One
of the optically identified cross-matched galaxies (MGTH\_J021759.3-041634) was missed by both of these approaches. It was the galaxy with the highest surface brightness of the cross-matched galaxies, and thus it is
reasonable that it was the one that was missed in the target selection. For completeness, we added it to the list of galaxies for surface brightness photometry.

\subsection{Surface brightness photometry}
\label{sec:surf_bright_phot}
In total, there are 34 galaxies for which surface brightness photometry was performed. Ultra-deep data are preferred over deep data for photometry whenever available. 
The surface brightness photometry was done in two steps. In the first step, we conducted isophotal fitting on the $i$ band image, with which we determined the geometry of the stellar disk, as detailed in \citetads{2024A&A...692A.217S}. In the second step we extracted surface brightness profiles in $g, r,$ and $i$ bands following the procedure described in appendix A of \citetads{2023A&A...670A..92M} and fit these profiles with a Sérsic function to derive the surface brightness parameters.
These steps are described briefly below.

The isophotal fitting step uses masked $i$ band images smoothed with a Gaussian with a full width half maximum (FWHM) of 0.94\arcsec.  A two-stage isophotal fitting was undertaken.
The first stage was used to constrain the center of the galaxy, and in the second stage, the ellipticity and position angle were constrained. The final ellipticity and position angle values were taken as the median from the outer extents of the galaxies, at surface brightnesses between 24 and 27 mag arcsec$^{-2}$. Optical inclinations were derived with
\begin{equation}
    \cos^2 i = \frac{(1-\epsilon)^2 - q_0^2}{1-q_0^2},
\end{equation}
where $\epsilon$ denotes the ellipticity and $q_0$ is the intrinsic thickness of the stellar disk. We took $q_0=0.3$ as a common value for the dwarf irregular galaxies \citepads[e.g.,][]{2010MNRAS.406L..65S} and applied it to almost all galaxies in the sample, except J022429.6-044037, for which the ellipticity was above 0.7 (hence giving an unphysical inclination). For this case, we set the inclination to $85_{-5}^{+4}$\dg.
The final geometric parameters of all galaxies are presented in Appendix \ref{app:phot}
(Table \ref{tab:global_phot}).

Using the derived geometric parameters, we created a set of ellipses of increasing semimajor axes with the same geometry and with widths equal to the FWHM of the PSF. In each ellipse, we took the mean value of the corresponding pixels in image units and corrected for inclination by multiplying with $\cos i$. The profile was extracted until $S/N = 1$ in a given ellipse. 
We converted image units (counts) to magnitudes using\footnote{\url{https://hsc-release.mtk.nao.ac.jp/doc/index.php/faq__pdr3}}: 
\begin{equation}
    m\ \text{[mag]} = -2.5\log_{10} ( \Sigma\, \text{counts}) + 27.
\end{equation}

To characterize the obtained surface brightness profiles, we fit them with a Sérsic function. We independently fit profiles in all three bands ($X \in [g,r,i]$) for each galaxy, and we derived the following quantities: the central surface brightness ($\mu_{0,X}$), Sérsic index ($n_X$), effective radius ($r_{\text{eff},X}$), mean effective surface brightness ($\overline{\mu}_{\mathrm{eff,X}}$), and the apparent magnitude ($m_X$). Details
on the fitting and error propagation are available in \citetads{2024A&A...692A.217S}. The extracted profiles and the accompanying Sérsic fits are presented in Appendix \ref{app:phot}.

As some of the galaxies measured in this section are at very low redshift (with three galaxies below $z \lesssim 0.02$), we updated from using the cosmological distances based solely on redshift to using distances based on a flow model, where local motions of galaxies are taken into account when computing a distance from a velocity.  We used Cosmic Flows 4 \citepads{2020AJ....159...67K,2024NatAs...8.1610V} to determine the luminosity distances to all the sources. Errors on distances were obtained using the prescription from \citetads{2025A&A...696A.185H}. We calculated the angular diameter distance from the luminosity distance as
\begin{equation}
    D_{A} = D_{L} / (1+z)^2.
\end{equation}
The various distances are collected in Table \ref{tab:global_phot}, while the photometric values for
each band are
reported in Tables \ref{tab:g_phot}, \ref{tab:r_phot} and \ref{tab:i_phot}, along with Galactic extinction corrections from  \citetads{2011ApJ...737..103S}\footnote{
The HSC employs a very similar photometric system \citepads{2018PASJ...70...66K} as the Sloan Digital Sky Survey (SDSS) \citepads{1996AJ....111.1748F}.} obtained through the NASA/IPAC Extragalactic Database (NED).

\subsection{Newly identified LSBGs}

Of the 34 galaxies for which we performed surface brightness photometry, 22 meet the LSBG
requirements of G18 (\mueffg$ > 24.3$ and $r_{\text{eff,g}} = 2.5-14$\arcsec). Of the 12 measured galaxies that did not meet the G18 criteria, one was too bright, two were too small and too bright, and nine were too small.
These \hi-identified LSBGs are listed in Table \ref{tab:sb_phot} with their key photometric values.
Reassuringly, the three galaxies identified in the cross-match with G18 and L23 are identified as LSBs by this definition.
Eleven of these sources were identified automatically, and eleven were only identified visually.
Thus, both the automatic and visual classification were important for identifying LSBGs.

\subsection{Newly identified UDGs}
With the available distance information, we were able to additionally classify the galaxies as UDGs or not based
on the definition $\overline{\mu}_{\mathrm{eff,g}} > 24$ mag arcsec$^{-2}$ and $R_{\text{eff},g} > 1.5$ kpc \citepads{2016A&A...590A..20V}. In this case
because we used the available distance information, we also
applied the correction for cosmological dimming to derive the intrinsic effective surface brightness
when classifying galaxies as UDGs.
This correction excluded two galaxies, leaving us with
26 UDGs, seven of which are not classified as LSBGs due to having angular sizes smaller than 2.5\arcsec. These galaxies are also listed in Table \ref{tab:sb_phot}. 
Of the three cross-matched galaxies, two are classified as UDGs.
While the angular size of MGTH\_J022138.1-052631 is 
found to be larger with the bespoke photometry of deeper HSC data,
it is still not large enough to be classified as a UDG.
 Of the 26, 15 were identified in the automated approach and 11 only via visual inspection,
 again demonstrating the utility of both approaches.

\begin{table*}
\centering
\caption{Surface brightness photometry of LSBGs and UDGs.}
\begin{tabular}{llllllllllll}
\hline \hline
Name & D$_{L, CF}$ & $\overline{\mu}_{\mathrm{eff,g}}$\tablefootmark{a} & e($\overline{\mu}_{\mathrm{eff,g}}$) & $r_{\mathrm{eff,g}}$ & e($r_{\mathrm{eff,g}}$) & $R_{\mathrm{eff,g}}$ & e($R_{\mathrm{eff,g}}$) & g-i\tablefootmark{a} & e(g-i) & LSBG & UDG \\
 & Mpc & arcsec$^{-2}$ &  & \arcsec &  & kpc &  &  &  &  &  \\
\hline
MGTH\_J021808.2-045217 & 104.0 & 25.71 & 0.01 & 2.56 & 0.07 & 1.23 & 0.13 & 0.40 & 0.00 & yes & no \\
MGTH\_J022138.1-052631 & 25.9 & 24.39 & 0.09 & 10.94 & 0.42 & 1.35 & 0.24 & 0.51 & 0.01 & yes & no \\
MGTH\_J022042.1-052115 & 30.1 & 24.45 & 0.04 & 5.02 & 0.28 & 0.72 & 0.13 & 0.59 & 0.03 & yes & no \\
MGTH\_J022051.9-045832 & 189.1 & 25.21 & 0.02 & 3.97 & 0.08 & 3.34 & 0.30 & 0.69 & 0.01 & yes & yes \\
MGTH\_J022429.6-044037 & 190.1 & 25.52 & 0.73 & 7.44 & 2.05 & 6.29 & 1.82 & 0.56 & 0.27 & yes & yes \\
MGTH\_J022400.9-044943 & 190.1 & 24.73 & 0.04 & 3.13 & 0.06 & 2.65 & 0.24 & 0.52 & 0.01 & yes & yes \\
MGTH\_J022358.3-050756 & 227.2 & 24.98 & 0.03 & 2.68 & 0.05 & 2.67 & 0.23 & 0.17 & 0.01 & yes & yes \\
MGTH\_J022357.0-051918 & 347.7 & 25.43 & 0.09 & 4.12 & 0.39 & 5.99 & 0.73 & 0.49 & 0.04 & yes & yes \\
MGTH\_J022350.6-045045 & 190.1 & 25.49 & 0.16 & 12.76 & 1.06 & 10.79 & 1.30 & 0.22 & 0.23 & yes & yes \\
MGTH\_J022332.2-044928 & 192.9 & 24.80 & 0.25 & 6.46 & 1.01 & 5.53 & 0.99 & 0.55 & 0.07 & yes & yes \\
MGTH\_J022117.8-045040 & 187.8 & 24.43 & 0.03 & 2.63 & 0.09 & 2.20 & 0.20 & 0.28 & 0.01 & yes & yes \\
MGTH\_J022443.4-045158 & 176.8 & 24.86 & 0.06 & 3.32 & 0.13 & 2.62 & 0.25 & 0.41 & 0.08 & yes & yes \\
MGTH\_J022045.0-050058 & 384.1 & 25.17 & 0.01 & 2.69 & 0.28 & 4.27 & 0.55 & 0.31 & 0.01 & yes & yes \\
MGTH\_J022522.7-045519 & 151.1 & 24.85 & 0.10 & 6.24 & 0.30 & 4.25 & 0.44 & 0.40 & 0.00 & yes & yes \\
MGTH\_J022026.3-045912 & 183.7 & 24.95 & 0.05 & 3.36 & 0.21 & 2.75 & 0.30 & 0.41 & 0.01 & yes & yes \\
MGTH\_J021848.1-045028 & 391.0 & 24.84 & 0.09 & 4.67 & 0.21 & 7.53 & 0.65 & 0.43 & 0.02 & yes & yes \\
MGTH\_J021818.0-043234 & 184.9 & 24.54 & 0.04 & 3.02 & 0.10 & 2.49 & 0.23 & 0.02 & 0.01 & yes & yes \\
MGTH\_J021759.3-041634 & 138.6 & 24.74 & 0.15 & 10.71 & 1.31 & 6.74 & 1.04 & 0.56 & 0.09 & yes & yes \\
MGTH\_J021724.3-043322 & 140.9 & 25.23 & 0.05 & 2.69 & 0.08 & 1.72 & 0.17 & 0.53 & 0.01 & yes & yes \\
MGTH\_J021645.4-042035 & 389.9 & 25.09 & 0.08 & 4.73 & 0.16 & 7.61 & 0.62 & 0.36 & 0.05 & yes & yes \\
MGTH\_J021642.0-044221 & 184.0 & 25.33 & 0.07 & 6.80 & 0.52 & 5.57 & 0.65 & 0.38 & 0.04 & yes & yes \\
MGTH\_J022029.4-041207 & 82.3 & 24.62 & 0.05 & 6.89 & 0.21 & 2.64 & 0.30 & 0.37 & 0.03 & yes & yes \\
MGTH\_J022056.9-045917 & 345.9 & 25.16 & 0.02 & 1.58 & 0.04 & 2.29 & 0.18 & 0.87 & 0.01 & no & yes \\
MGTH\_J022058.7-051320 & 291.4 & 24.58 & 0.03 & 1.43 & 0.05 & 1.78 & 0.15 & 0.87 & 0.02 & no & yes \\
MGTH\_J022110.7-050630 & 177.9 & 24.97 & 0.05 & 2.20 & 0.10 & 1.75 & 0.18 & 0.27 & 0.03 & no & yes \\
MGTH\_J022011.5-051047 & 225.9 & 25.22 & 0.01 & 1.96 & 0.03 & 1.94 & 0.16 & 0.41 & 0.01 & no & yes \\
MGTH\_J022338.7-042431 & 310.5 & 24.51 & 0.06 & 2.13 & 0.10 & 2.81 & 0.25 & 0.51 & 0.00 & no & yes \\
MGTH\_J021625.5-044228 & 183.9 & 24.32 & 0.04 & 2.20 & 0.10 & 1.80 & 0.18 & 0.38 & 0.01 & no & yes \\
MGTH\_J021534.4-042904 & 381.2 & 25.92 & 0.02 & 1.95 & 0.11 & 3.08 & 0.28 & 1.30 & 0.01 & no & yes \\
\hline
\end{tabular}
\tablefoottext{a}{Extinction-corrected}
\label{tab:sb_phot}
\end{table*}

\section{\hi\ content of optically identified LSBGs}\label{sec:hi_content}
While the optically identified LSBGs mostly fall in the outskirts
of the MIGHTEE ES footprint and do not have direct \hi\ detections, we can examine the known locations of the LSBGs in the MIGHTEE \hi\ data cubes to search for marginal detections or to put an upper limit on their \hi\ fluxes. 
Below we describe extraction of \hi\ spectra at the known locations of the optical LSBGs and interpret the (non-)detections.

\subsection{\hi\ spectra}
Before extracting \hi\ spectra to search for marginal detections or place upper limits on non-detections, it was important to first consider where it is meaningful to extract data from in terms of the overall sensitivity of the data, i.e., the primary beam response.
As can be seen in Figure \ref{fig:xmm}, \hi\ sources can be detected well beyond even the 30\% primary beam response level. However, this depends on the intrinsic strength of the signal, and the data cubes themselves extend very far out in the primary beam response. 
To enforce some minimum quality on our flux estimates, we extracted the \hi\ spectra only for sources within the 5\% limit of the primary beam response. 
For each pointing there are three spectral windows that cover the full redshift range. We applied this requirement to each of these frequency cubes separately.

We applied the primary beam correction to the cubes before extracting the spectra. If a source was covered by more than one pointing, we combined the spectra using a linear weighting scheme based on the primary beam response. In all cases, we extracted the spectra
using a 30\arcsec\ aperture, which is consistent with expected \hi\ source sizes.\footnote{
The three cross-matched sources in Section \ref{sec:crossmatch} have \hi\ sizes of 28\arcsec -- 55\arcsec based on the relation of \citetads{2022MNRAS.512.2697R}. 
}
Figure \ref{fig:hi_spec} shows the extracted \hi\ spectra
for all detections in G18 and L23 that fell within 5\% of the primary beam response in at least one spectral window.

\begin{figure}
    \centering
    \includegraphics[width=0.9\linewidth]{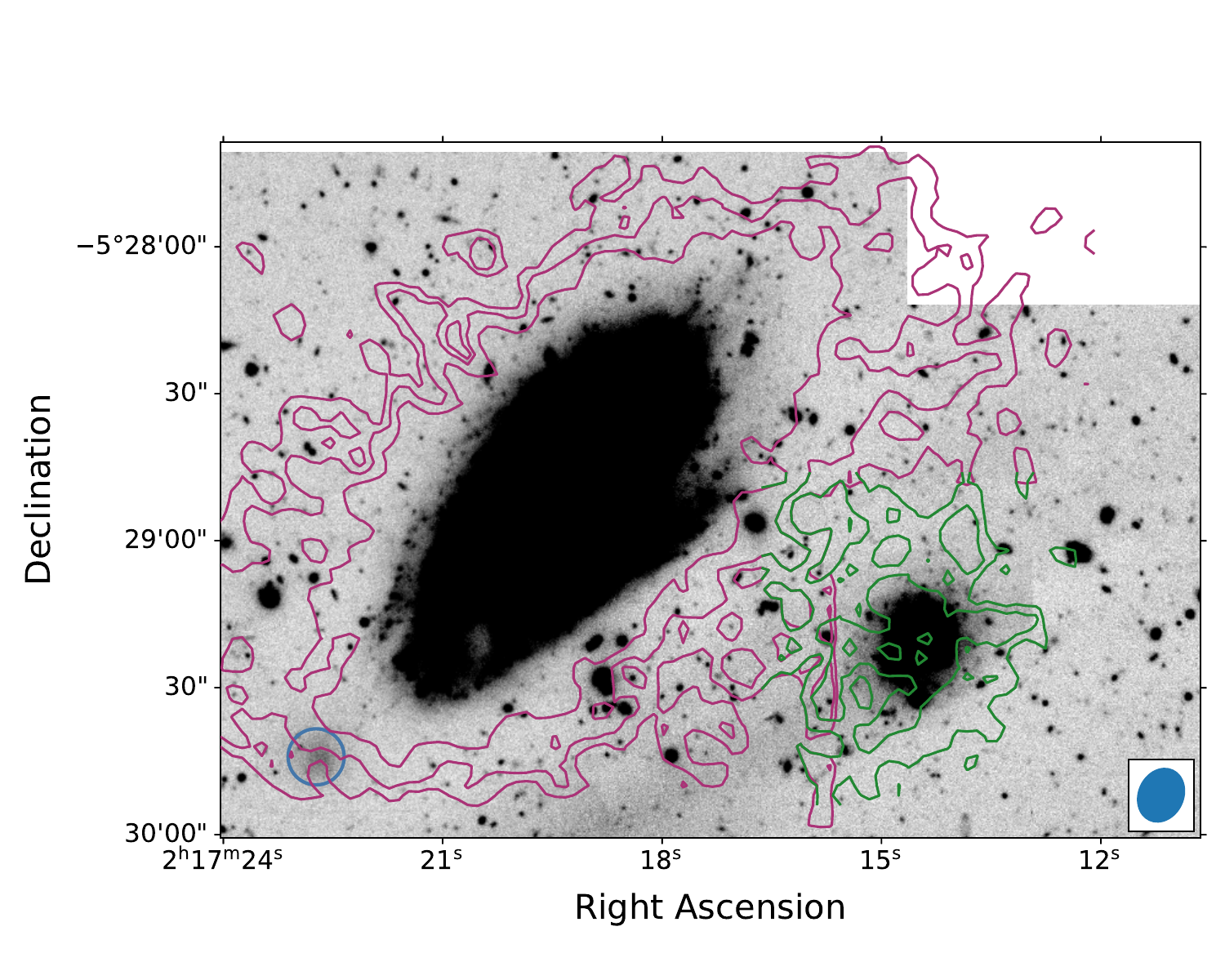}
    \caption{Optically detected LSBG, G327 (blue circle), near two \hi-detected galaxies. Red \hi\ contours are MGTH\_J021719.4-
052851 and green contours are MGTH\_J021714.5-052922. The levels are [2.5, 5, 10] $\times 10^{20}$ atoms cm$^{-2}$ for both. 
}
    \label{fig:g327}
\end{figure}

The three cross-matched sources (G267, L25657, and L86749) are clearly seen in these spectra. 
G327 also has a detection, but this optical detection lies next to a dwarf galaxy merging with a massive galaxy (both gas rich). As seen in Figure \ref{fig:g327}, the aperture for the spectrum for G327 clearly falls on top of the \hi\ emission for the larger galaxy in this system, MGTH\_J021719.4-052851. With the poor spectral resolution of the early science data, it is not possible to distinguish whether there may be distinct \hi\ emission arising from G327. 
We assumed the \hi\ emission is fully from MGTH\_J021719.4-052851 and used the rest of the spectrum for calculating an upper limit for G327.

No other sources have clear detections. A few sources have peaks around 4$\sigma$, but other sources have similar negative peaks. As a confirmation, we visually inspected the cubes for L5787, L45406, and G322. There was no evidence of emission for L45406 or G322. 
For L5787 (G315), there is emission at the frequency of the tentative detection ($\sim$1410 MHz); this is the very tip of the northern \hi\ arm of NGC\,895 
\citepads{2023MNRAS.521.5177N}. Given the poor spectral resolution of the MIGHTEE \hi\ ES data, it is not possible to say if there may be separate \hi\ emission at this spatial and spectral location associated with L5787 or if it is a happen-chance alignment of a background source. With the full science data at full spectral resolution, it will be interesting to explore this feature to determine if L5787 may be a faint satellite of NGC\,895 or a tidal feature. For now, we treat it as a non-detection. We report the standard deviations of the spectra in Table \ref{tab:hi_limits}.

\begin{table*}
\centering
\caption{\hi\ properties of optically selected galaxies.}
\begin{tabular}{llllllll}
\hline \hline
Name & RA & Dec & $z_{host}$ & $z_{min}$ & $z_{max}$ & rms & \mhi/\mstar \\
 &  &  &  &  &  & mJy &  \\
\hline
G262 & 36.46018 & -4.4021 &  & 0.0011 & 0.084 & 0.66 & $<18.10$ \\
G265 & 36.37673 & -5.04815 &  & 0.0011 & 0.084 & 0.25 & $< 8.36$ \\
G266 & 36.3633 & -4.96164 &  & 0.0011 & 0.084 & 0.22 & $< 5.47$ \\
G267 & 34.17509 & -4.70562 &  & 0.0011 & 0.084 & 0.11 & 9.51 \\
G268 & 35.18785 & -4.53235 &  & 0.0011 & 0.084 & 0.13 & $< 1.50$ \\
G304 & 36.52391 & -5.25631 &  & 0.0011 & 0.084 & 0.82 & $<19.18$ \\
G308 & 36.58562 & -5.33722 &  & 0.0293 & 0.084 & 1.64 & $<13.41$ \\
G309 & 36.33909 & -5.5587 &  & 0.0293 & 0.084 & 1.43 & $<41.79$ \\
G318 & 35.35374 & -5.73281 &  & 0.068 & 0.084 & 2.09 & $< 6.56$ \\
G319 & 35.44907 & -5.65036 &  & 0.0011 & 0.084 & 1.17 & $< 9.11$ \\
G322 & 33.84203 & -5.27461 &  & 0.0011 & 0.084 & 0.58 & $<10.97$ \\
G327 & 34.34471 & -5.49559 &  & 0.0011 & 0.084 & 0.35 & $< 7.73$ \\
L21295 & 35.343 & -5.423 & 0.032 & 0.0011 & 0.084 & 0.33 & $< 5.54$ \\
L25657 & 34.497 & -4.276 & 0.035 & 0.0011 & 0.084 & 0.26 & 3.52 \\
L30385 & 35.333 & -5.516 & 0.032 & 0.0011 & 0.084 & 0.49 & $<23.22$ \\
L35220 & 34.514 & -4.306 & 0.034 & 0.0011 & 0.084 & 0.24 & $<30.57$ \\
L45406 & 35.35 & -5.359 & 0.032 & 0.0011 & 0.084 & 0.25 & $<20.19$ \\
L55931 & 35.315 & -5.539 & 0.032 & 0.0011 & 0.084 & 0.53 & $< 5.49$ \\
L5787 & 35.361 & -5.434 & 0.032 & 0.0011 & 0.084 & 0.36 & $<34.81$ \\
L86749 & 35.409 & -5.442 & 0.032 & 0.0011 & 0.084 & 0.47 & 9.72 \\
\hline
\end{tabular}
\label{tab:hi_limits}
\end{table*}

\subsection{Gas content}
Without a direct detection, we do not have a redshift and
thus could not place direct limits on the \hi\ mass, a distance-dependent quantity. However, the spectra can still be used to place upper limits on the integrated flux, and we can examine \mhi/\mstar, a distance-independent quantity.\footnote{
Note this is not strictly true as stellar masses require a k-correction that depends on redshift. However, for the redshift range considered here, this correction is small, and thus this statement is approximately true.
}

We took the standard deviation of the spectra, assumed galaxies would occupy two channels (88 \kms\ or 417.968 kHz),\footnote{
The assumed width is important, as a narrower linewidth is easier to detect. We adopted two channels as a reasonable value for the range of galaxies we were focused on. The three cross-matched galaxies have \w50\ values of 46--99 \kms, which are consistent with our assumed width. 
} and calculated a 5$\sigma$ upper limit for the integrated line flux, which is used to derive the upper limit for the \hi\ mass.
We applied the approach of L23 for calculating stellar masses,
using the average of the ($g-r$) and ($g-i$) mass-to-light ratios from \citetads{2013MNRAS.430.2715I},
to the G18 galaxies. 

The \mhi/\mstar values are reported in Table \ref{tab:hi_limits}.
For the three cross-matches, these are measured values.\footnote{We note that we used the color-based mass-to-light stellar masses for consistency (see Section \ref{sec:phot_diff}), but the cross-matched sources do have SED-based stellar masses available.} For all other sources, the values are 5$\sigma$ upper limits. In general, the \mhi/\mstar\ upper limits are quite high, so the lack of \hi\ detections in these galaxies is not surprising. When examining Figure \ref{fig:xmm}, it is clear that the optically identified LSBGs tend to lie at the edges of the ES footprint, where sensitivity is low. The full science data in XMM-LSS will include expanded spatial coverage with full sensitivity, so hopefully they will provide either detections or more constraining upper limits.

\section{Discussion}\label{sec:discussion}

\subsection{Differences between photometry approaches}\label{sec:phot_diff}
To compare the \hi- and optically selected galaxies, 
it is important to understand the differences in the photometric approaches and the impact they might have.
In the G18 catalog, the surface brightness distribution was modeled as a two-dimensional PSF-convolved Sérsic function in the $i$ band. In the L23 catalog, the surface brightness distribution was modeled using a Spergel profile convolved with the PSF. In this work, one-dimensional surface brightness profiles sampled by the FWHM of the PSF were fit with a Sérsic function.

To assess the impact of the different photometric approaches, we compared the
photometric properties of the three cross-matched galaxies in Table \ref{tab:comp_phot}. While there are only three sources, we observed that two of them are generally consistent between the different approaches to surface brightness photometry. The third source is measured to be much brighter and more extended in this work. This source is also the galaxy that had an erroneous host assignment in L23. Visual inspection of the images did not reveal an obvious explanation for the difference in extent (and total magnitude), but it showed that our photometry encompasses the full source extent. 

\begin{table}[]
    \centering
    \caption{Comparison of photometric measurements.}
    \resizebox{0.48\textwidth}{!}{%
    \begin{tabular}{llll}
    \hline \hline
    Names & Property\tablefootmark{a} & This work & G18/L23 \\
    \hline
    J021642.0-044221/ &  g-i & 0.42$\pm$0.04 & 0.43$\pm$ 0.24 \\
    G267 & $m_i$ & 18.85$\pm$0.03 & 18.89 $\pm$0.24 \\
         & \re\ (\arcsec)\tablefootmark{b} &  6.44$\pm$0.42 & 6.27$\pm$0.79 \\
         & $\overline{\mu}_{\mathrm{eff,g}}$\tablefootmark{c} & 25.18$\pm$0.25 & 25.30$\pm$0.24\\
    \hline
    J021759.3-041634/ &  g-i & 0.59$\pm$0.09 & 0.66 $\pm$0.05\\
    L25657 & $m_i$ & 17.07$\pm$0.04 & 17.02 $\pm$ 0.07\\
        & \re\ (\arcsec)\tablefootmark{b} & 10.21 $\pm$ 0.76 & 8.52 $\pm$ 0.35 \\
        & $\overline{\mu}_{\mathrm{eff,g}}$\tablefootmark{c} & 24.66 $\pm$0.18  & 24.33$\pm$0.05 \\
    \hline
    J022138.1-052631/ &  g-i & 0.54$\pm$0.01 & 0.43$\pm$0.05 \\
    L86749 & $m_i$ & 16.73$\pm$0.01 &  18.99$\pm$0.08\\
        & \re\ (\arcsec)\tablefootmark{b} &11.17$\pm$0.48 & 5.37$\pm$0.31 \\
        & $\overline{\mu}_{\mathrm{eff,g}}$\tablefootmark{c} & 24.24$\pm$0.15 & 25.06$\pm$0.06  \\
    \hline
    \end{tabular}}
    \label{tab:comp_phot}
    \tablefoottext{a}{Not corrected for extinction}
    \tablefoottext{b}{Measured in the $i$ band}
    \tablefoottext{c}{$\overline{\mu}_{\mathrm{eff,g}} = m_g + 2.5\log_{10}{(2\pi \text{\re}^2)}$}
\end{table}

\begin{figure}
    \centering
    \includegraphics[width=0.9\linewidth]{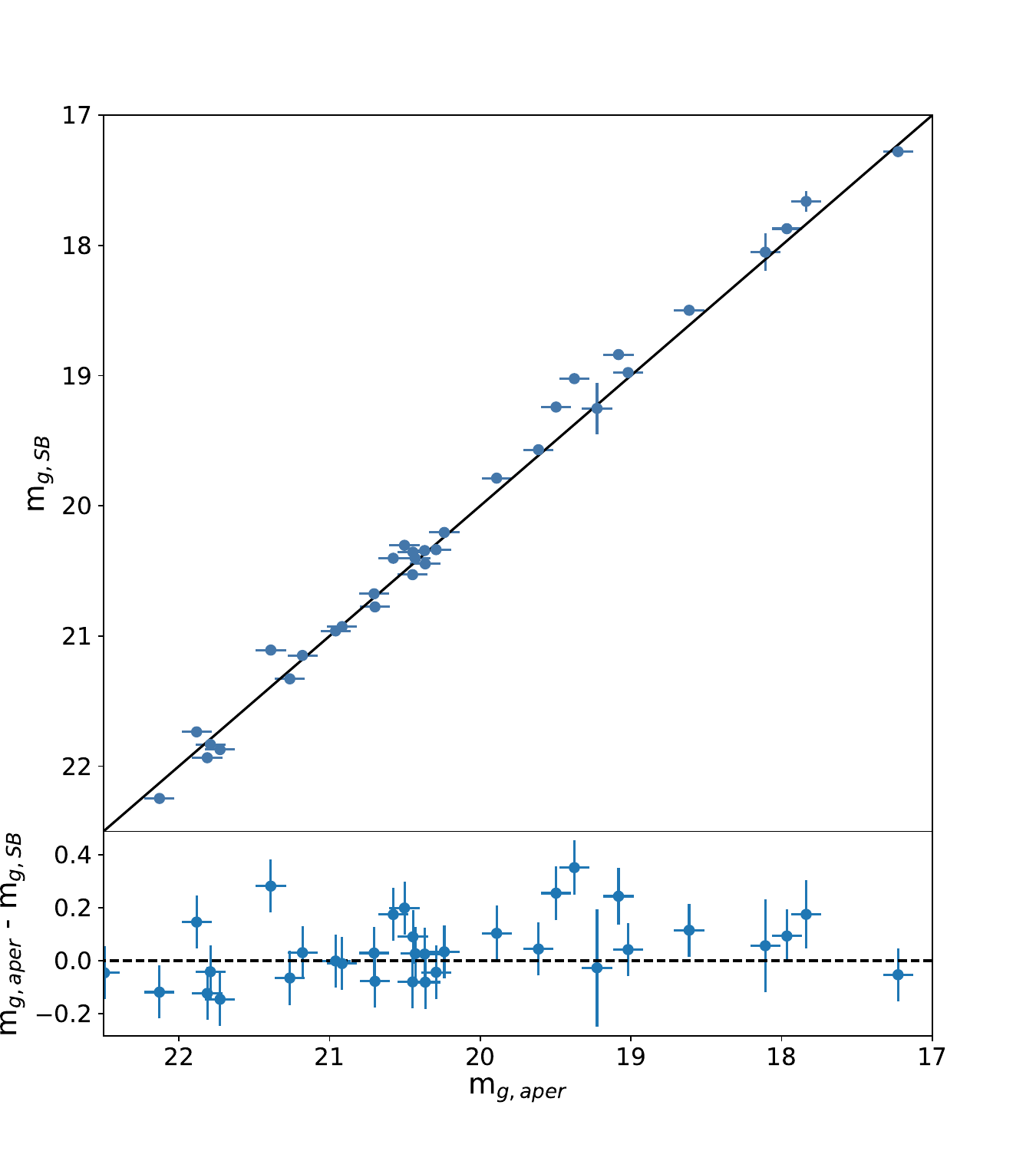}
    \caption{Comparison of apparent (non-extinction corrected) $g$ band magnitudes derived through aperture and surface brightness photometry. \textit{Upper:} Magnitudes derived through surface brightness photometry (y-axis) plotted against those derived from aperture photometry (x-axis). \textit{Lower:} Differences between the two magnitudes plotted against the apparenet magnitude from aperture photometry.}
    \label{fig:phot_comp}
\end{figure}

Furthermore, as the aperture magnitudes from the MIGHTEE ES catalog were used for the SED-based stellar masses, ensuring the consistency of the aperture magnitudes with our derived photometry allows us to use the stellar masses in the MIGHTEE ES catalog. Figure \ref{fig:phot_comp} presents a comparison of the two sets of photometry, showing that the two approaches result in consistent measurements of the total magnitude. However, when comparing the SED-based stellar masses to those derived using mass-to-light (ML) ratios (as for L23), we found a systematic offset with the ML-based stellar masses that is $\sim0.3$ dex higher than the SED-based stellar masses. The
SED-based stellar masses are generally considered more robust, while the ML-based stellar masses require less observational data. The systematic offset that we found is indicative of the underlying systematics present in stellar mass estimates of all types. Given this, we always explicitly state what stellar mass estimates we use throughout the rest of this section. 

\subsection{HI versus optically identified populations}

The optically and \hi\ identified LSBGs and UDGs in this work are mostly distinct populations, with only three galaxies in common. Here we explore their 
properties to better understand the differences between the two populations.
Throughout this section, for the three galaxies detected in both \hi\ and optical searches, we plot them with their properties as measured in this work and the original optical measurements from either L23 or G18. We highlight these three galaxies so it is clear that they are the common galaxies.

\begin{figure}
    \centering
    \includegraphics[width=0.9\linewidth]{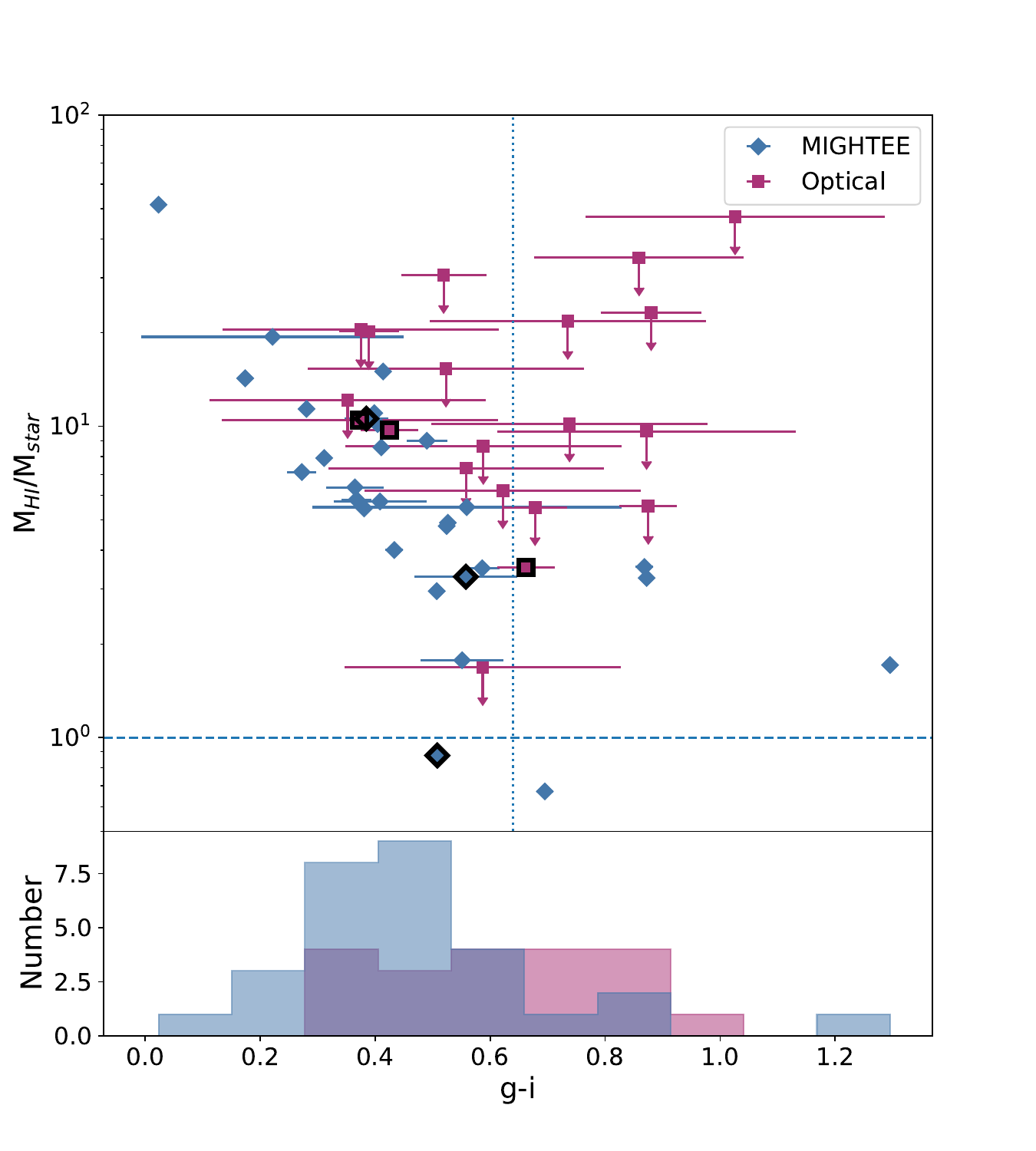}
    \caption{Colors of the optically and \hi-selected galaxies. \textit{Upper:} g-i color vs. \mhi/\mstar. Downward arrows indicate upper limits. Symbols highlighted in black indicate the cross-matched galaxies that are plotted twice. The vertical dotted line at 0.64 indicates the cut between "red" and "blue" used in G18. The dashed horizontal line indicates a \mhi/\mstar\ value of one.
    \textit{Lower:} Histogram of the color distribution of both the MIGHTEE-detected (blue) and optically detected galaxies (pink).}
    \label{fig:color_gasfrac}
\end{figure}

Figure \ref{fig:color_gasfrac} shows a comparison of the gas fraction 
and color (g-i) of the
MIGHTEE-detected LSBGs and UDGs of this work in relation to the LSBGs of G18 and the UDGs and UPGs of L23. In this case, we used stellar masses based on the
same ML ratio for both samples to enable the closest comparison.
The \hi-detected galaxies of this work are predominantly blue, as expected for \hi-selected
galaxies.
Some of the optically identified galaxies exhibit similar blue colors, and thus it may 
be initially surprising that not more of the optically identified galaxies were detected in the MIGHTEE \hi\ data. However, the upper limits on the \hi\ content of the optically identified galaxies are not very constraining, with upper limits on \mhi/\mstar ranging from $\sim5-34$. Thus, the bluer optically identified galaxies may still host significant amounts of \hi. It is also worth noting the large gas fractions of the LSBGs  that do have \hi\ detections -- only two galaxies in total have more mass in stars than \hi.

\begin{figure}
    \centering
    \includegraphics[width=0.9\linewidth]{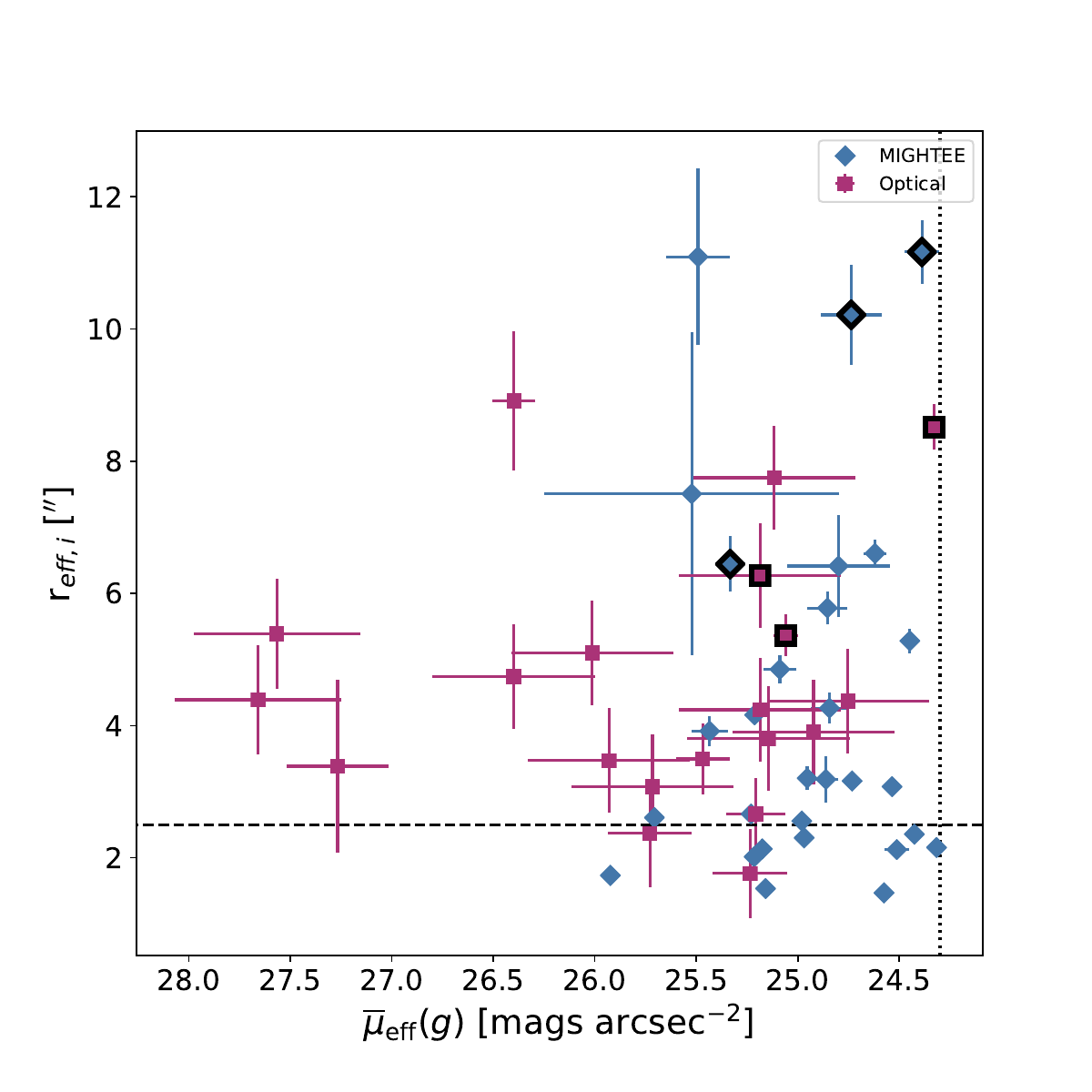}
    \caption{\mueffg\ color vs. \re. Symbols highlighted in black indicate galaxies that are identified by both MIGHTEE and the works of G18 and L23, and they are plotted twice (once for each measurement). The vertical dotted line and dashed horizontal lines indicate the LSBG cuts of G18 for surface brightness and effective radius.
    }
    \label{fig:mueff_reff}
\end{figure}

Given the non-stringent \hi\ gas fraction limits,
it is understandable why the optically identified blue LSBGs are not detected in MIGHTEE. More puzzling is that the MIGHTEE-detected LSBGs and UDGs are not
identified in the optical searches. For L23, this may be partially explained by the requirement of
a host galaxy for an associated redshift, as the \hi-selected population from MIGHTEE may not have the host galaxies required by L23. In addition, L23 used a fainter effective surface brightness cut for their definition of a UDG that would exclude some of the UDGs in our definition. For comparison to G18, we looked at the two essential metrics, \re\ and \mueffg, as seen in Figure \ref{fig:mueff_reff}. The MIGHTEE UDGs that are not also LSBGs would be excluded from G18 due to their angular size being smaller than the minimum used in their selection. 
Additionally, it is possible that the automated source finding and characterization of G18 and L23 would result in larger sources being shredded if the emission was patchy, as is often the case for \hi-rich galaxies.
Overall, there is not one clear reason that the MIGHTEE LSBGs were missed in the optical searches, but it highlights the utility of targeted photometry based on \hi\ detections for identifying LSBGs.

\subsection{Impacts of survey footprint and redshift limits}
As discussed above, very few of the optically identified LSBGs are detected in the MIGHTEE ES data. This can be mostly explained by the footprint of the MIGHTEE ES data. As seen in Figure \ref{fig:xmm}, most of the optical detections fall in the outskirts of the ES data, with all except two beyond the 50\% response level and almost half beyond the 30\% response level. However, it is also worth considering the impact of redshift limits, as the detection of \hi\ requires that the galaxy has a redshift that falls within the frequency coverage of the observation. As seen in Table \ref{tab:hi_limits}, the redshifts of the associated hosts for the galaxies from L23 all fall within the limits of the MIGHTEE ES data ($z=0-0.084$); thus these would be detected if the observations were sensitive enough, assuming the host associations are accurate. The galaxies from G18 do not have redshifts available; however, \citetads{2022ApJ...933..150G} used a statistical approach to estimate that $\sim$50\% of the galaxies are at $z<0.05$, with essentially no galaxies beyond $z=0.15$. The MIGHTEE redshift limits extend to $z=0.084$, and thus we might reasonably expect the majority of G18 galaxies to fall within the observed frequency range and to be detected if there were sufficient sensitivity.

\subsection{The baryonic Tully-Fisher relation}
\begin{figure}
    \centering
    \includegraphics[width=0.9\linewidth]{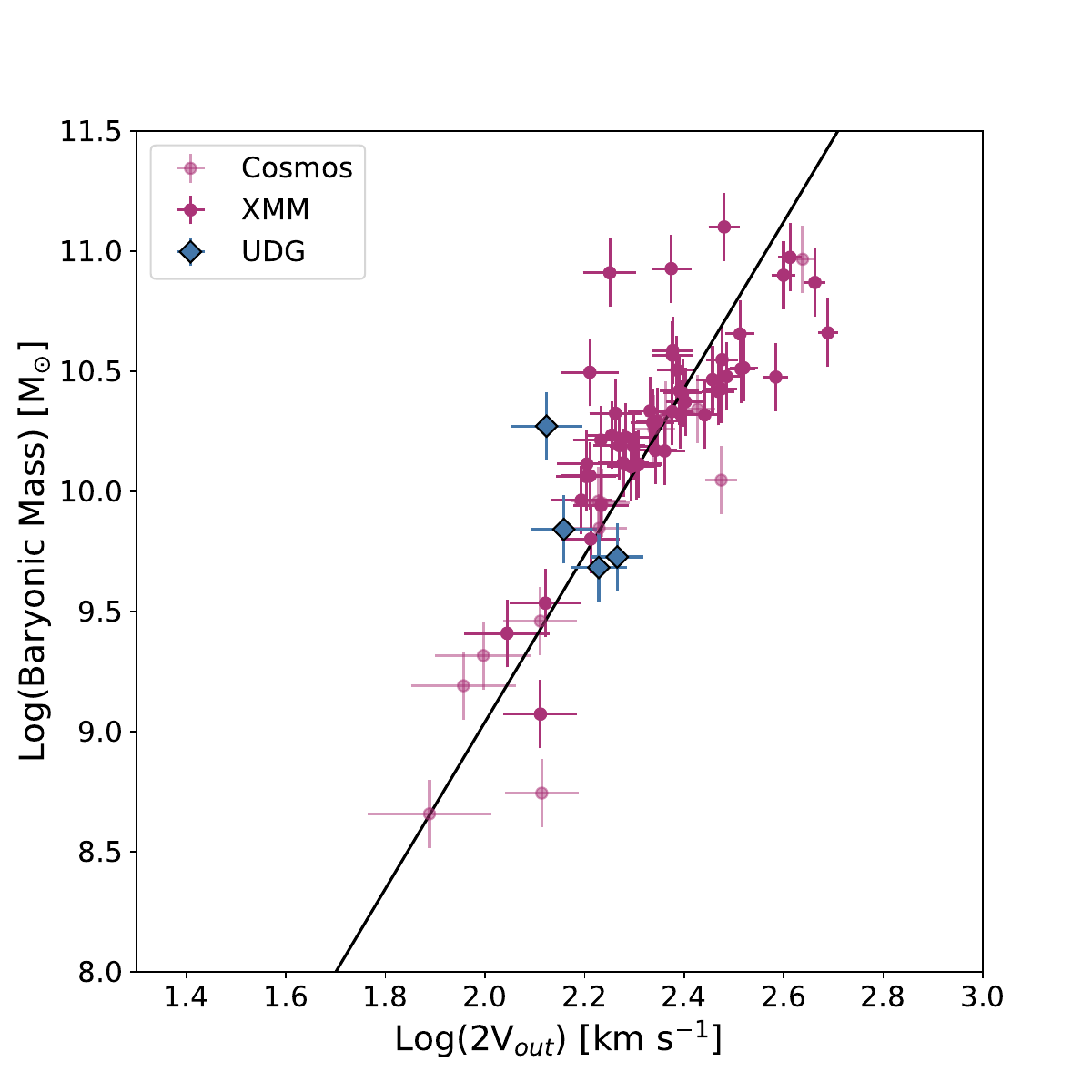}
    \caption{
    Baryonic mass vs. outermost rotational velocity. MIGHTEE galaxies are shown with magenta circles, with galaxies from the COSMOS field (not searched for LSBG/UDGs) in a lighter magenta. The four galaxies of this work (meeting both LSBG and UDG classification criteria) are shown as blue diamonds.
    The derived bTFr from  \citetads{2021MNRAS.508.1195P} is shown by the black line.
    }
    \label{fig:btfr}
\end{figure}

The baryonic Tully-Fisher relation (bTFr) relates total baryonic mass to the rotational velocity of a galaxy. It is one of the tightest scaling relations known, with extremely low intrinsic scatter \citep[e.g.,][]{2016ApJ...816L..14L,2016A&A...593A..39P,2018MNRAS.474.4366P,2019MNRAS.484.3267L}. Recently, some gas-rich UDGs have been proposed to be outliers to this relation 
\citepads[e.g.,][]{2019ApJ...883L..33M,2020MNRAS.495.3636M,2023ApJ...947L...9H,
2024A&A...692A.217S} due to
having lower rotation velocities than expected for their baryonic mass, potentially revealing new insights into galaxy formation \citepads[e.g.,][]{2020MNRAS.495.3636M,2024A&A...689A.344M,2025MNRAS.538...60A}. \citetads{2021MNRAS.508.1195P} studied the bTFr for all MIGHTEE ES galaxies where robust
kinematic models could be derived.
Figure \ref{fig:btfr} shows the bTFr using the outermost rotational velocity, V$_{out}$, from \citetads{2021MNRAS.508.1195P}, and the LSBGs and UDGs identified in this work are highlighted. There are only four galaxies from this work that have the kinematic information required to be shown in this plot, and all four are classified as both UDG and LSBG.  Three of these galaxies are fully consistent with this relation, while one is potentially an outlier. Interestingly, this potential outlier (J022350.6-045045) has one of the largest physical sizes, with an effective radius of 10.79 kpc, 
although such a large size could be indicative of disruption or non-equilibrium \citepads[e.g.,][]{2015ApJ...809L..21M,2017A&A...608A.142V}. However, there are other galaxies in the XMM-LSS field that were not identified in this work as an LSBG or UDG that show similar offsets from the bTFr.
The relatively poor spectral resolution of the early science data naturally biases us toward sources that are spectrally wide and hence more likely to have higher rotational velocities for a given inclination. This limits the exploration of UDGs that might be offset from the bTFr toward lower rotational velocities. 
Deriving updated and additional rotation velocities with the MIGHTEE full science data at the full velocity resolution will help reveal how the UDG and LSBG populations as a whole connect to the bTFr.

\subsection{LSBGs in an \hi-identified population}
\begin{figure}
    \centering
    \includegraphics[width=0.9\linewidth]{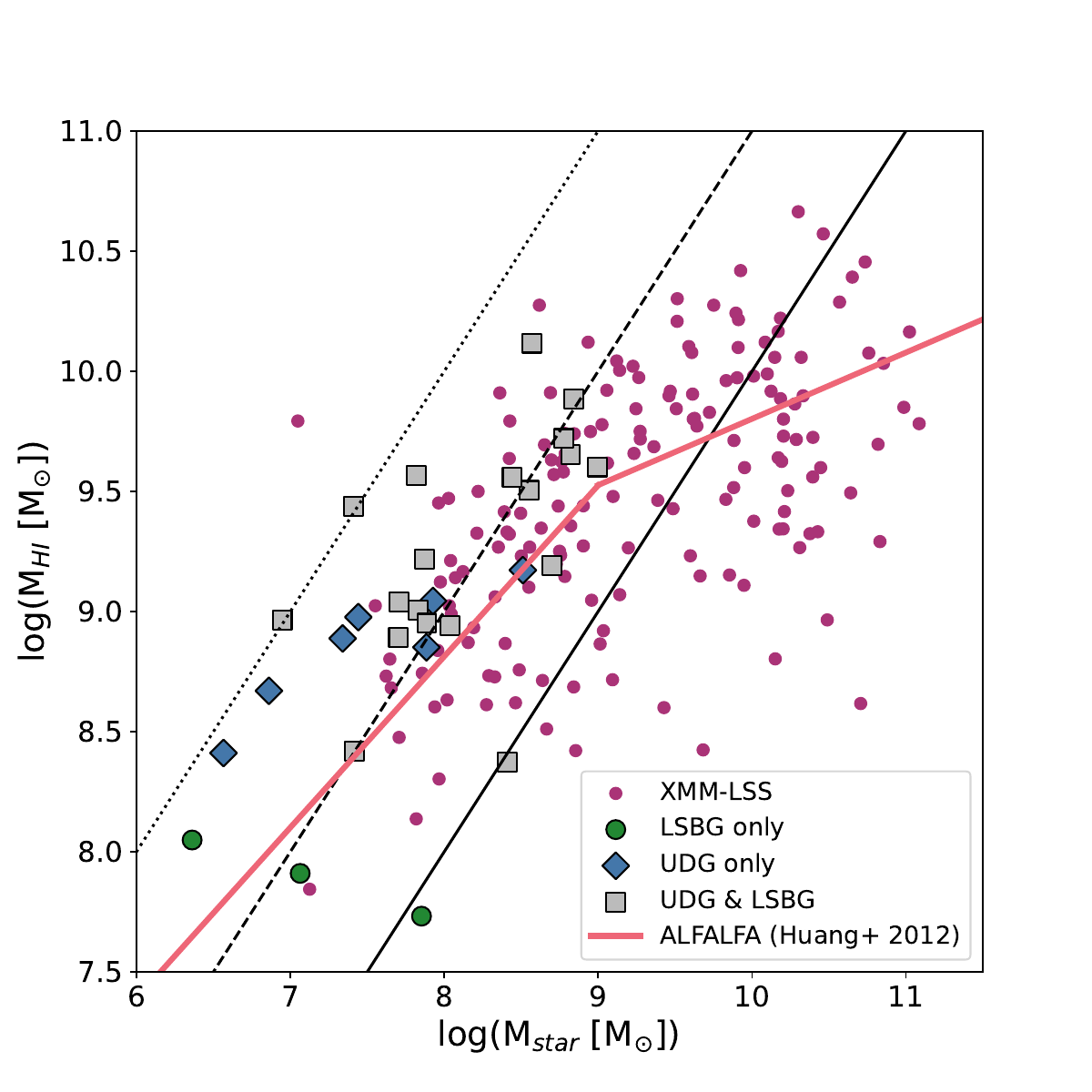}
    \caption{\mhi\ vs. \mstar\ for all MIGHTEE galaxies in XMM-LSS. The UDGs and LSBGs are indicated, and the diagonal solid, dashed, and dotted lines indicate \mhi\ to \mstar\ ratios of 1, 10, and 100. All stellar masses are derived from SED estimates. Galaxies from this work that meet only the LSBG definition (green circles), UDG definition (blue diamonds), or both (gray squares) are indicated separately. The solid red line indicates the relation from the ALFALFA survey \citepads{2012ApJ...756..113H}.}
    \label{fig:mhi_mstar}
\end{figure}

The LSBGs and UDGs we identified here appear to be very gas rich, as seen in Figure \ref{fig:color_gasfrac}. We examined this further with Figure \ref{fig:mhi_mstar}, looking at \mhi\ versus \mstar\ for all MIGHTEE detections in XMM-LSS and using the SED-based stellar masses derived in a consistent manner for all MIGHTEE detections. The LSBGs and UDGs are clearly more gas rich than the full MIGHTEE sample \citepads{2021A&A...646A..35M}.  
Thus, for a given stellar mass, lower surface brightness galaxies tend to have more mass in \hi\ than those with a higher surface brightness. As seen in the previous section, there seems to be no trend with surface brightness in our sample for galaxies to lie on or fall off the bTFr (for the subset with adequate kinematics in the ES data).
This implies that for a given rotation velocity (or dark matter halo mass), galaxies of different surface brightnesses have similar baryonic masses. Thus, the more gas-rich nature of the LSBGs and UDGs indicates that they have not turned their gas into stars as efficiently, in agreement with the arguments put forth by \citetads{2022ApJ...941...11K}.

It is also worthwhile to briefly compare the UDGs and LSBGs of this work to other \hi-selected samples. \citetads{2017ApJ...842..133L} provided the first catalog of \hi-selected UDGs from the ALFALFA \hi\ survey, with their broad sample having $\overline{\mu}_{\mathrm{eff,r}} > 24$ mag arcsec$^{-2}$. They found that the \hi-selected UDGs were gas rich compared to the general ALFALFA population, akin to how we find that our UDGs and LSBGs are gas rich compared to other MIGHTEE detections. As MIGHTEE \hi\ detections have a similar \mhi\ and \mstar\ compared to ALFALFA \citepads[][]{2021A&A...646A..35M}, our UDGs and LSBGs are gas rich in comparison to that larger reference population as well, as can also be seen in Figure \ref{fig:mhi_mstar}. 
More recently, \citetads{2025PASA...42...87O} identified LSBGs in the WALLABY pilot survey data. Their sample appears to have a similar behavior for \mhi-\mstar\ when compared to ALFALFA (see their figure 5). However, they used a less stringent requirement of \mueffg $>23$ mag arcsec$^{-2}$ for their definition of LSBGs. Thus, it could be that it is only at a lower surface brightnesses (e.g., $\gtrsim 24$ mag arcsec$^{-2}$) that there is a distinction in the \hi\ mass versus stellar mass for galaxies. Exploring gas-richness in different regimes of surface brightness is an interesting avenue for future exploration.

\section{Conclusions}\label{sec:conclusions}

In this work, we have used the MIGHTEE early science \hi\ data in the XMM-LSS field to understand the LSBG population, including UDGs. This was done by searching for new LSBGs missed in the previous optical searches and studying the gas content of all known LSBGs within the footprint. 

Cross-matching of the \hi\ detections and optically identified LSBGs from G18 and L23 revealed three sources in common. The MIGHTEE \hi\ detection provides direct redshifts for these galaxies. One redshift is completely new, and a previous redshift assignment is updated.

Thirty-four MIGHTEE galaxies were chosen as potential LSBG candidates. Dedicated surface brightness photometry revealed that 29 met either the LSBG definition of G18 (\mueffg $>24.3$ mag arcsec$^{-2}$ and 2.5\arcsec < \re\ <14\arcsec; 22 galaxies) and/or the adopted UDG definition (\mueffg $>24$ mag arcsec$^{-2}$ and $R_{\text{eff},g} >$ 1.5 kpc; 26 galaxies). Furthermore, 20 optically identified galaxies from G18 and L23 fell within the bounds of the MIGHTEE ES footprint (i.e., within 5\% of the primary beam response of at least one field). We performed a targeted search for \hi\ emission at the locations of these optically selected galaxies but identified no new \hi\ emission beyond the three cross-matched sources. The upper limits on the gas fractions for these galaxies are not constraining, as the optically identified galaxies fall mostly on the edges of the MIGHTEE ES footprint, with reduced sensitivity.

Comparison between the \hi- and optically selected samples showed that \hi-detected galaxies tend toward bluer colors on average, but in general the \hi-selected sample overlaps
with the optically selected sample in color, surface brightness, and size.
The lack of identification of the \hi-selected galaxies in the optical catalogs is likely due to a combination of different selection cuts in the optical catalogs (e.g., minimum sizes and requiring host galaxies) combined with a targeted surface brightness photometry approach for patchy LSB sources instead of automated source detection and modeling.

The LSBGs and UDGs of MIGHTEE were examined and compared to the full MIGHTEE population.
Only four have available kinematics from \citetads{2021MNRAS.508.1195P} , three of which show no clear offset of the full population from the baryonic Tully-Fisher relation. Our analysis, however, is biased toward galaxies with larger rotational velocities due to the poor spectral resolution of the data.
Further studies with more LSBGs and careful derivation of the kinematics are therefore warranted. This will be especially important when the MIGHTEE data at full spectral resolution are available
in this region. 
The MIGHTEE LSBGs are also clearly gas rich compared to the full population of MIGHTEE detections, pointing to a scenario where their LSB nature arises from not efficiently converting gas to stars. A comparison to other \hi\ samples, namely ALFALFA \citepads{2017ApJ...842..133L} and WALLABY \citepads{2025PASA...42...87O}, indicated that a surface brightnesses lower than $\sim$24 mag arcsec$^{-2}$ may be where gas fractions start to significantly change.

The optical and \hi\ identification both play important roles in identifying LSBGs. Here we have shown samples from the same footprint, where LSBGs identified via their stellar versus gaseous content are similar in number but with very little overlap, showing the utility of using both tracers to understand the full population. As both optical and \hi\ surveys increase in depth, we may expect the overlap to grow between the two tracers but for both to always have valuable roles to play: optical for identifying gas-poor LSBGs and \hi\ for identifying ``too shy to shine'' galaxies that have minimal stellar components.

\begin{acknowledgements}
We thank 
the anonymous referee for their helpful comments, and  J. M. van der Hulst for insightful discussion.
PEMP is funded by the Dutch Research Council (NWO) through the Veni grant VI.Veni.222.364.
MB gratefully acknowledges financial support from the Flemish Fund for Scientific Research (FWO-Vlaanderen) and the South African National Research Foundation (NRF) under their Bilateral Scientific Cooperation program (grant G0G0420N), and from the Belgian Science Policy Office (BELSPO), under grant BL/02/SA12 (GALSIMAS). MJJ acknowledges the support of the STFC consolidated grant [ST/S000488/1] and [ST/W000903/1] , a UKRI Frontiers and  the Hintze Family Charitable Foundation through the Oxford Hintze Centre for Astrophysical Surveys. GS acknowledge funding from IRMIA++, SARAO (Grant no.: 97882), and the European Union’s Horizon 2020 research and innovation programme under the Marie Skłodowska‑Curie grant agreement No 101147719.

The MeerKAT telescope is operated by the South African Radio Astronomy Observatory, which is a facility of the National Research Foundation, an agency of the Department of Science, Technology and Innovation. We acknowledge the use of the ilifu cloud computing facility – www.ilifu.ac.za, a partnership between the University of Cape Town, the University of the Western Cape, Stellenbosch University, Sol Plaatje University and the Cape Peninsula University of Technology. The ilifu facility is supported by contributions from the Inter-University Institute for Data Intensive Astronomy (IDIA – a partnership between the University of Cape Town, the University of Pretoria and the University of the Western Cape), the Computational Biology division at UCT and the Data Intensive Research Initiative of South Africa (DIRISA). This work made use of the CARTA (Cube Analysis and Rendering Tool for Astronomy) software (DOI 10.5281/zenodo.3377984 – https://cartavis.github.io). The Hyper Suprime-Cam (HSC) collaboration includes the astronomical communities of Japan and Taiwan, and Princeton University. The HSC instrumentation and software were developed by the National Astronomical Observatory of Japan (NAOJ), the Kavli Institute for the Physics and Mathematics of the Universe (Kavli IPMU), the University of Tokyo, the High Energy Accelerator Research Organization (KEK), the Academia Sinica Institute for Astronomy and Astrophysics in Taiwan (ASIAA), and Princeton University. Funding was contributed by the FIRST program from Japanese Cabinet Office, the Ministry of Education, Culture, Sports, Science and Technology (MEXT), the Japan Society for the Promotion of Science (JSPS), Japan Science and Technology Agency (JST), the Toray Science Foundation, NAOJ, Kavli IPMU, KEK, ASIAA, and Princeton University.  This work made use of Astropy (\url{http://www.astropy.org}) a community-developed core Python package and an ecosystem of tools and resources for astronomy \citep{astropy:2013, astropy:2018, astropy:2022}.

\end{acknowledgements}

\bibliographystyle{aa}
\bibliography{refs}

\clearpage
\raggedbottom
\onecolumn

\appendix

\setcounter{table}{0}
\setcounter{figure}{0}
\renewcommand{\thetable}{\Alph{section}.\arabic{table}}
\renewcommand{\thefigure}{\Alph{section}.\arabic{figure}}

\section{Surface brightness photometry}\label{app:phot}

\begin{figure}
    \centering
    \includegraphics[width = 0.88 \textwidth]{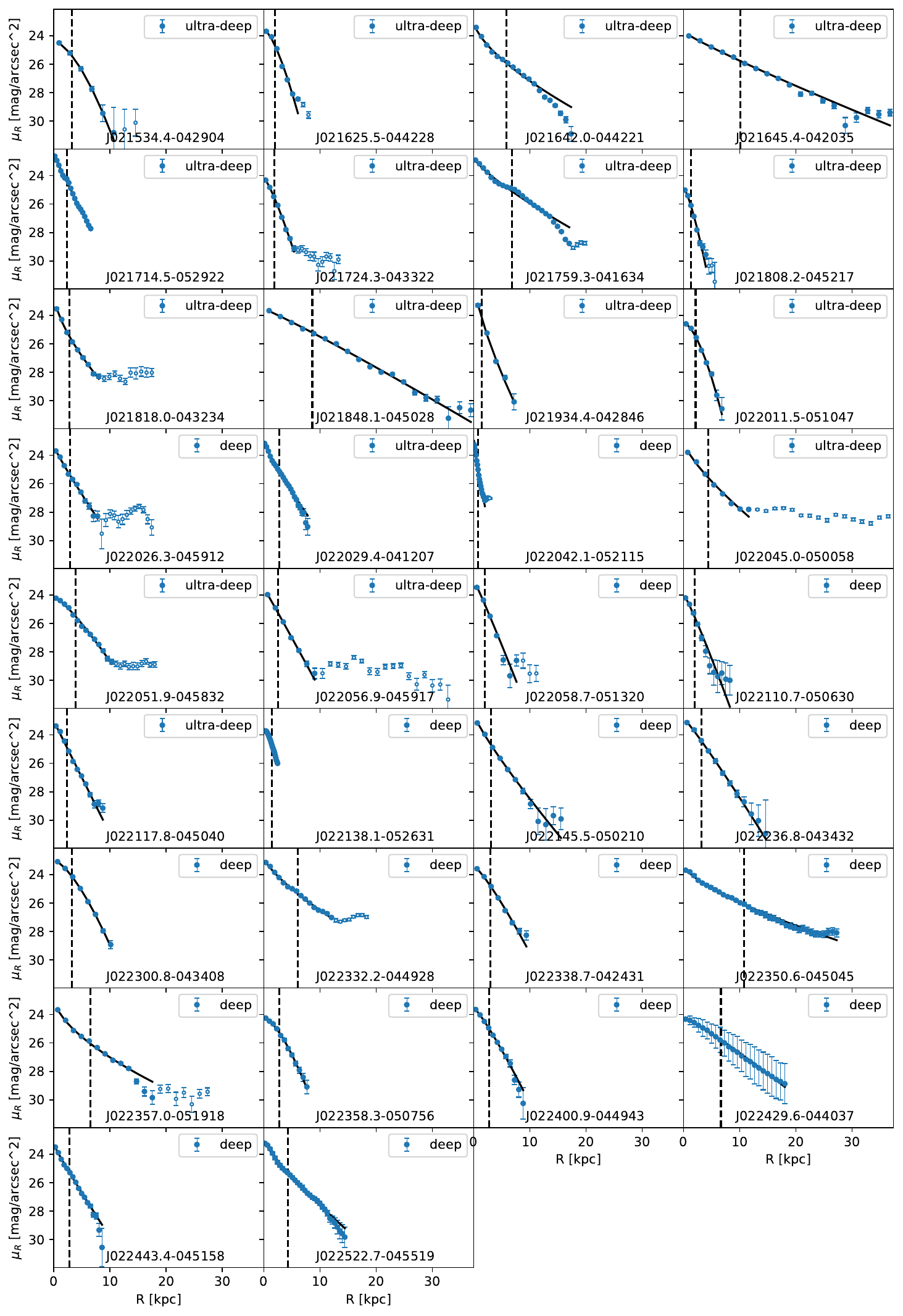}\\
    \caption{Surface brightness profiles of all 34 candidate LSBGs selected from the MIGHTEE ES \hi\ catalog. Blue circles with error bars correspond to the galaxy profiles extracted from the images, solid black lines show the resulting Sérsic fit, and vertical dashed black lines denote the effective radius obtained from the fit.}
    \label{fig:surf_bright_profs}
\end{figure}

This appendix presents the surface brightness photometry values for
all galaxies that were measured. Table \ref{tab:global_phot} presents the global values for all galaxies, including their name,
 the depth of the data, the various distances to the galaxy, and
the global optical morphology, and Tables \ref{tab:g_phot}, \ref{tab:r_phot}
and \ref{tab:i_phot} present the photometry for the $g$, $r$ and $i$ bands, not corrected for Galactic extinction. In addition, we show the surface brightness profiles and Sérsic profile fits from which all these properties were derived in Figure \ref{fig:surf_bright_profs}.

\bigskip
\begin{minipage}{\textwidth}
\captionof{table}{Global properties of all galaxies with surface brightness photometry.}
\label{tab:global_phot}
\centering
\begin{tabular}{lllrrrrrrrr}
\hline \hline
Name & Data & z & D$_{L,z}$\tablefootmark{a} & D$_{L,CF}$\tablefootmark{b} & e(D$_{L,CF}$) & D$_{AD,CF}$\tablefootmark{c} & $i$ & e($i$) & PA & e(PA) \\
MGTH\_ &  &  & Mpc &  &  &  & \dg & \dg & \dg & \dg \\
\hline
J021534.4-042904 & Ultra & 0.082098 & 350.2 & 381.2 & 28.5 & 325.6 & 72 & 2 & 19.4 & 0.8 \\
J021625.5-044228 & Ultra & 0.042752 & 186.1 & 183.9 & 16.1 & 169.2 & 29 & 11 & -134.2 & 29.2 \\
J021642.0-044221 & Ultra & 0.042752 & 186.1 & 184.0 & 16.1 & 169.2 & 41 & 10 & -58.4 & 21.4 \\
J021645.4-042035 & Ultra & 0.083651 & 356.6 & 389.9 & 29.0 & 332.0 & 54 & 11 & -9.9 & 30.6 \\
J021714.5-052922 & Ultra & 0.017464 & 77.0 & 69.5 & 8.2 & 67.1 & 32 & 8 & -33.4 & 48.4 \\
J021724.3-043322 & Ultra & 0.034075 & 149.0 & 140.9 & 13.3 & 131.8 & 56 & 5 & -17.7 & 3.9 \\
J021759.3-041634 & Ultra & 0.033398 & 146.1 & 138.6 & 13.1 & 129.8 & 45 & 8 & -76.0 & 15.6 \\
J021808.2-045217 & Ultra & 0.025756 & 113.1 & 104.0 & 10.7 & 98.8 & 55 & 3 & 1.9 & 3.9 \\
J021818.0-043234 & Ultra & 0.042912 & 186.8 & 184.9 & 16.2 & 170.0 & 62 & 5 & 12.7 & 4.5 \\
J021848.1-045028 & Ultra & 0.083866 & 357.4 & 391.0 & 29.0 & 332.9 & 50 & 14 & 77.8 & 28.1 \\
J021934.4-042846 & Ultra & 0.069497 & 298.4 & 315.2 & 24.4 & 275.5 & 46 & 2 & -44.6 & 0.8 \\
J022011.5-051047 & Ultra & 0.05195 & 225.1 & 225.9 & 18.8 & 204.1 & 40 & 3 & 55.8 & 5.9 \\
J022026.3-045912 & Deep & 0.042591 & 185.4 & 183.7 & 16.1 & 169.0 & 51 & 6 & -87.0 & 3.5 \\
J022029.4-041207 & Ultra & 0.020365 & 89.7 & 82.3 & 9.1 & 79.0 & 66 & 2 & -46.5 & 8.2 \\
J022042.1-052115 & Deep & 0.007359 & 32.6 & 30.1 & 4.9 & 29.6 & 46 & 8 & -69.3 & 4.0 \\
J022045.0-050058 & Ultra & 0.082615 & 352.3 & 384.1 & 28.6 & 327.7 & 27 & 4 & -12.5 & 19.5 \\
J022051.9-045832 & Ultra & 0.043873 & 190.9 & 189.1 & 16.5 & 173.5 & 60 & 3 & -28.9 & 6.1 \\
J022056.9-045917 & Ultra & 0.075589 & 323.5 & 345.9 & 26.3 & 299.0 & 32 & 7 & 81.7 & 16.7 \\
J022058.7-051320 & Deep & 0.064971 & 279.6 & 291.4 & 22.9 & 256.9 & 52 & 3 & 25.0 & 2.9 \\
J022110.7-050630 & Deep & 0.041315 & 180.0 & 177.9 & 15.7 & 164.1 & 70 & 2 & -45.1 & 1.4 \\
J022117.8-045040 & Ultra & 0.043552 & 189.5 & 187.8 & 16.4 & 172.5 & 54 & 4 & -73.3 & 3.0 \\
J022138.1-052631 & Deep & 0.006912 & 30.6 & 25.9 & 4.6 & 25.5 & 66 & 4 & 49.6 & 2.5 \\
J022145.5-050210 & Deep & 0.073552 & 315.2 & 335.4 & 25.7 & 291.0 & 57 & 5 & -85.4 & 3.7 \\
J022236.8-043432 & Deep & 0.069496 & 298.4 & 314.9 & 24.4 & 275.3 & 75 & 3 & 28.6 & 0.8 \\
J022300.8-043408 & Deep & 0.073381 & 314.4 & 334.5 & 25.6 & 290.3 & 43 & 9 & 46.6 & 11.9 \\
J022332.2-044928 & Deep & 0.044673 & 194.3 & 192.9 & 16.7 & 176.7 & 55 & 10 & 28.1 & 23.2 \\
J022338.7-042431 & Deep & 0.068655 & 294.9 & 310.5 & 24.1 & 271.9 & 45 & 7 & 59.2 & 12.1 \\
J022350.6-045045 & Deep & 0.044031 & 191.5 & 190.1 & 16.5 & 174.4 & 71 & 5 & -18.2 & 15.2 \\
J022357.0-051918 & Deep & 0.07593 & 324.9 & 347.7 & 26.4 & 300.3 & 30 & 12 & -78.0 & 24.0 \\
J022358.3-050756 & Deep & 0.052276 & 226.4 & 227.2 & 18.9 & 205.2 & 59 & 4 & -45.9 & 5.3 \\
J022400.9-044943 & Deep & 0.044032 & 191.5 & 190.1 & 16.5 & 174.4 & 63 & 3 & -85.0 & 5.2 \\
J022429.6-044037 & Deep & 0.044031 & 191.5 & 190.1 & 16.5 & 174.4 & 85 & 5 & 20.6 & 0.5 \\
J022443.4-045158 & Deep & 0.040994 & 178.6 & 176.8 & 15.6 & 163.2 & 51 & 5 & 26.1 & 14.6 \\
J022522.7-045519 & Deep & 0.036864 & 160.9 & 151.1 & 13.9 & 140.5 & 74 & 4 & -29.2 & 1.5 \\
\hline
\end{tabular}
\tablefoottext{a}{Cosmological luminosity distance}
\tablefoottext{b}{Flow model based luminosity distance}
\tablefoottext{c}{Angular diameter distance from flow model}
\end{minipage}

\begin{minipage}{\textwidth}
\captionof{table}{Surface brightness photometry values in the $g$ band.}
\label{tab:g_phot}
\centering
\footnotesize
\begin{tabular}{llllllllllllll}
\hline \hline
Name & $A_g$ & $m_g$ & e($m_{g}$) & $\mu_{0,g}$ & e($\mu_{0,g}$) & $\overline{\mu}_{eff,g}$ & e($\overline{\mu}_{eff,g}$) & $n_g$ & e($n_{g}$) & $r_{\text{eff},g}$ & e($r_{\text{eff},g}$) & $R_{\text{eff},g}$ & e($R_{\text{eff},g}$) \\
MGTH\_ & mag & \multicolumn{2}{c}{mag} & 
\multicolumn{2}{c}{mag arcsec$^{-2}$} & 
\multicolumn{2}{c}{mag arcsec$^{-2}$} & 
 & & \multicolumn{2}{c}{arcsec} & 
 \multicolumn{2}{c}{kpc} \\ \hline
J021534.4-042904 & 0.067 & 22.54 & 0.01 & 25.35 & 0.24 & 25.99 & 0.02 & 0.70 & 0.09 & 1.95 & 0.11 & 3.39 & 0.54 \\
J021625.5-044228 & 0.067 & 20.68 & 0.00 & 23.80 & 0.22 & 24.38 & 0.04 & 0.67 & 0.08 & 2.20 & 0.10 & 1.98 & 0.31 \\
J021642.0-044221 & 0.065 & 19.24 & 0.02 & 23.50 & 0.38 & 25.40 & 0.07 & 1.44 & 0.16 & 6.80 & 0.52 & 6.12 & 1.03 \\
J021645.4-042035 & 0.069 & 19.79 & 0.04 & 24.01 & 0.17 & 25.16 & 0.08 & 1.01 & 0.07 & 4.73 & 0.16 & 8.38 & 1.29 \\
J021714.5-052922 & 0.077 & 17.87 & 0.00 & 23.00 & 0.14 & 24.00 & 0.04 & 0.93 & 0.05 & 6.71 & 0.16 & 2.28 & 0.35 \\
J021724.3-043322 & 0.066 & 21.15 & 0.01 & 24.58 & 0.08 & 25.30 & 0.05 & 0.75 & 0.02 & 2.69 & 0.08 & 1.90 & 0.29 \\
J021759.3-041634 & 0.069 & 17.66 & 0.08 & 23.29 & 0.42 & 24.81 & 0.15 & 1.23 & 0.17 & 10.71 & 1.31 & 7.41 & 1.44 \\
J021808.2-045217 & 0.067 & 21.74 & 0.00 & 25.11 & 0.12 & 25.77 & 0.01 & 0.72 & 0.05 & 2.56 & 0.07 & 1.33 & 0.20 \\
J021818.0-043234 & 0.066 & 20.21 & 0.01 & 22.71 & 0.22 & 24.60 & 0.04 & 1.44 & 0.09 & 3.02 & 0.10 & 2.74 & 0.42 \\
J021848.1-045028 & 0.067 & 19.57 & 0.01 & 24.03 & 0.14 & 24.91 & 0.09 & 0.85 & 0.04 & 4.67 & 0.21 & 8.30 & 1.30 \\
J021934.4-042846 & 0.073 & 21.94 & 0.01 & 22.10 & 0.51 & 23.64 & 0.01 & 1.24 & 0.22 & 0.87 & 0.07 & 1.27 & 0.22 \\
J022011.5-051047 & 0.071 & 21.83 & 0.00 & 24.83 & 0.07 & 25.29 & 0.01 & 0.58 & 0.02 & 1.96 & 0.03 & 2.15 & 0.32 \\
J022026.3-045912 & 0.073 & 20.40 & 0.00 & 23.70 & 0.27 & 25.03 & 0.05 & 1.12 & 0.11 & 3.36 & 0.21 & 3.01 & 0.49 \\
J022029.4-041207 & 0.066 & 18.50 & 0.01 & 23.48 & 0.14 & 24.69 & 0.05 & 1.05 & 0.05 & 6.89 & 0.21 & 2.84 & 0.43 \\
J022042.1-052115 & 0.075 & 19.02 & 0.02 & 22.97 & 0.36 & 24.52 & 0.04 & 1.25 & 0.15 & 5.02 & 0.28 & 0.74 & 0.12 \\
J022045.0-050058 & 0.075 & 21.11 & 0.01 & 23.41 & 0.58 & 25.25 & 0.01 & 1.41 & 0.24 & 2.69 & 0.28 & 4.70 & 0.86 \\
J022051.9-045832 & 0.078 & 20.30 & 0.01 & 24.45 & 0.10 & 25.29 & 0.02 & 0.83 & 0.04 & 3.97 & 0.08 & 3.68 & 0.56 \\
J022056.9-045917 & 0.077 & 22.25 & 0.00 & 24.30 & 0.11 & 25.24 & 0.02 & 0.89 & 0.05 & 1.58 & 0.04 & 2.51 & 0.38 \\
J022058.7-051320 & 0.073 & 21.87 & 0.01 & 24.00 & 0.19 & 24.65 & 0.03 & 0.71 & 0.08 & 1.43 & 0.05 & 1.93 & 0.30 \\
J022110.7-050630 & 0.069 & 21.33 & 0.02 & 24.30 & 0.23 & 25.04 & 0.05 & 0.77 & 0.09 & 2.20 & 0.10 & 1.89 & 0.30 \\
J022117.8-045040 & 0.072 & 20.41 & 0.01 & 23.22 & 0.20 & 24.50 & 0.03 & 1.09 & 0.08 & 2.63 & 0.09 & 2.42 & 0.37 \\
J022138.1-052631 & 0.082 & 17.28 & 0.01 & 24.01 & 0.10 & 24.47 & 0.09 & 0.58 & 0.02 & 10.94 & 0.42 & 1.51 & 0.23 \\
J022145.5-050210 & 0.073 & 20.78 & 0.01 & 23.21 & 0.18 & 24.33 & 0.06 & 0.99 & 0.08 & 2.05 & 0.06 & 3.15 & 0.48 \\
J022236.8-043432 & 0.081 & 20.53 & 0.00 & 23.55 & 0.09 & 24.36 & 0.06 & 0.82 & 0.03 & 2.33 & 0.07 & 3.38 & 0.52 \\
J022300.8-043408 & 0.089 & 20.34 & 0.02 & 23.34 & 0.35 & 23.83 & 0.13 & 0.60 & 0.13 & 1.99 & 0.15 & 3.06 & 0.51 \\
J022332.2-044928 & 0.084 & 18.84 & 0.04 & 23.26 & 0.41 & 24.88 & 0.25 & 1.29 & 0.13 & 6.46 & 1.01 & 6.11 & 1.33 \\
J022338.7-042431 & 0.087 & 20.96 & 0.00 & 23.85 & 0.21 & 24.60 & 0.06 & 0.78 & 0.08 & 2.13 & 0.10 & 3.05 & 0.48 \\
J022350.6-045045 & 0.086 & 18.05 & 0.15 & 23.71 & 0.39 & 25.58 & 0.16 & 1.42 & 0.16 & 12.76 & 1.06 & 11.92 & 2.05 \\
J022357.0-051918 & 0.08 & 20.45 & 0.02 & 23.81 & 0.44 & 25.51 & 0.09 & 1.33 & 0.18 & 4.12 & 0.39 & 6.57 & 1.17 \\
J022358.3-050756 & 0.085 & 20.93 & 0.01 & 24.52 & 0.06 & 25.07 & 0.03 & 0.64 & 0.02 & 2.68 & 0.05 & 2.97 & 0.45 \\
J022400.9-044943 & 0.087 & 20.34 & 0.00 & 24.06 & 0.06 & 24.82 & 0.04 & 0.78 & 0.02 & 3.13 & 0.06 & 2.93 & 0.44 \\
J022429.6-044037 & 0.082 & 19.25 & 0.20 & 24.82 & 0.77 & 25.60 & 0.73 & 0.80 & 0.06 & 7.44 & 2.05 & 6.94 & 2.18 \\
J022443.4-045158 & 0.094 & 20.36 & 0.02 & 23.90 & 0.16 & 24.96 & 0.06 & 0.96 & 0.06 & 3.32 & 0.13 & 2.82 & 0.44 \\
J022522.7-045519 & 0.093 & 18.98 & 0.00 & 23.46 & 0.15 & 24.95 & 0.10 & 1.21 & 0.04 & 6.24 & 0.30 & 4.76 & 0.75 \\
\hline
\end{tabular}
\end{minipage}

\begin{minipage}{\textwidth}
\captionof{table}{Surface brightness photometry values in the $r$ band.}
\label{tab:r_phot}
\centering
\footnotesize
\begin{tabular}{llllllllllllll}
\hline \hline
Name & $A_r$ & $m_r$ & e($m_{r}$) & $\mu_{0,r}$ & e($\mu_{0,r}$) & $\overline{\mu}_{eff,r}$ & e($\overline{\mu}_{eff,r}$) & $n_r$ & e($n_{r}$) & $r_{\text{eff},r}$ & e($r_{\text{eff},r}$) & $R_{\text{eff},r}$ & e($R_{\text{eff},r}$) \\
MGTH\_ & mag & \multicolumn{2}{c}{mag} & 
\multicolumn{2}{c}{mag arcsec$^{-2}$} & 
\multicolumn{2}{c}{mag arcsec$^{-2}$} & 
 & & \multicolumn{2}{c}{arcsec} & 
 \multicolumn{2}{c}{kpc} \\ \hline
J021534.4-042904 & 0.047 & 21.70 & 0.01 & 24.35 & 0.05 & 24.88 & 0.03 & 0.63 & 0.02 & 1.72 & 0.02 & 2.99 & 0.45 \\
J021625.5-044228 & 0.046 & 20.42 & 0.00 & 23.52 & 0.22 & 24.13 & 0.04 & 0.69 & 0.09 & 2.21 & 0.10 & 1.99 & 0.31 \\
J021642.0-044221 & 0.045 & 18.95 & 0.02 & 23.01 & 0.41 & 25.04 & 0.07 & 1.51 & 0.18 & 6.58 & 0.47 & 5.92 & 0.98 \\
J021645.4-042035 & 0.048 & 19.39 & 0.04 & 23.72 & 0.18 & 25.03 & 0.09 & 1.11 & 0.06 & 5.35 & 0.21 & 9.49 & 1.47 \\
J021714.5-052922 & 0.053 & 17.55 & 0.00 & 22.49 & 0.14 & 23.71 & 0.04 & 1.06 & 0.06 & 6.83 & 0.17 & 2.33 & 0.35 \\
J021724.3-043322 & 0.046 & 20.76 & 0.00 & 24.14 & 0.10 & 24.94 & 0.05 & 0.81 & 0.03 & 2.74 & 0.07 & 1.94 & 0.30 \\
J021759.3-041634 & 0.048 & 17.26 & 0.08 & 22.72 & 0.49 & 24.31 & 0.15 & 1.27 & 0.20 & 10.22 & 1.24 & 7.07 & 1.37 \\
J021808.2-045217 & 0.046 & 21.47 & 0.01 & 24.85 & 0.16 & 25.56 & 0.01 & 0.75 & 0.06 & 2.62 & 0.09 & 1.36 & 0.21 \\
J021818.0-043234 & 0.045 & 20.14 & 0.01 & 22.62 & 0.30 & 24.61 & 0.05 & 1.49 & 0.13 & 3.12 & 0.14 & 2.83 & 0.44 \\
J021848.1-045028 & 0.046 & 19.26 & 0.00 & 23.54 & 0.16 & 24.54 & 0.10 & 0.93 & 0.05 & 4.56 & 0.21 & 8.10 & 1.27 \\
J021934.4-042846 & 0.05 & 21.50 & 0.01 & 21.61 & 0.84 & 23.37 & 0.02 & 1.36 & 0.36 & 0.94 & 0.12 & 1.37 & 0.27 \\
J022011.5-051047 & 0.049 & 21.53 & 0.01 & 24.49 & 0.09 & 25.00 & 0.01 & 0.62 & 0.03 & 1.97 & 0.04 & 2.17 & 0.33 \\
J022026.3-045912 & 0.05 & 20.13 & 0.00 & 23.37 & 0.20 & 24.65 & 0.05 & 1.09 & 0.08 & 3.19 & 0.17 & 2.86 & 0.45 \\
J022029.4-041207 & 0.046 & 18.16 & 0.02 & 23.00 & 0.13 & 24.39 & 0.04 & 1.16 & 0.05 & 7.03 & 0.23 & 2.89 & 0.44 \\
J022042.1-052115 & 0.052 & 18.54 & 0.02 & 22.71 & 0.28 & 24.24 & 0.04 & 1.24 & 0.12 & 5.52 & 0.27 & 0.82 & 0.13 \\
J022045.0-050058 & 0.052 & 20.89 & 0.01 & 23.09 & 0.45 & 24.76 & 0.01 & 1.31 & 0.19 & 2.36 & 0.19 & 4.13 & 0.70 \\
J022051.9-045832 & 0.054 & 19.79 & 0.01 & 24.09 & 0.11 & 24.90 & 0.01 & 0.81 & 0.04 & 4.20 & 0.09 & 3.90 & 0.59 \\
J022056.9-045917 & 0.053 & 21.67 & 0.01 & 23.50 & 0.14 & 24.58 & 0.03 & 0.98 & 0.06 & 1.53 & 0.04 & 2.43 & 0.37 \\
J022058.7-051320 & 0.051 & 21.25 & 0.02 & 22.93 & 0.45 & 23.96 & 0.01 & 0.94 & 0.19 & 1.39 & 0.10 & 1.87 & 0.31 \\
J022110.7-050630 & 0.048 & 21.10 & 0.02 & 23.93 & 0.24 & 24.88 & 0.05 & 0.90 & 0.09 & 2.27 & 0.13 & 1.95 & 0.31 \\
J022117.8-045040 & 0.05 & 20.17 & 0.00 & 23.04 & 0.17 & 24.16 & 0.04 & 1.00 & 0.07 & 2.52 & 0.08 & 2.32 & 0.36 \\
J022138.1-052631 & 0.057 & 16.90 & 0.01 & 23.65 & 0.10 & 24.13 & 0.09 & 0.60 & 0.02 & 11.18 & 0.43 & 1.55 & 0.24 \\
J022145.5-050210 & 0.05 & 20.55 & 0.01 & 22.56 & 0.18 & 23.95 & 0.07 & 1.16 & 0.08 & 1.91 & 0.07 & 2.95 & 0.46 \\
J022236.8-043432 & 0.056 & 20.17 & 0.00 & 22.82 & 0.10 & 23.81 & 0.06 & 0.92 & 0.03 & 2.12 & 0.06 & 3.08 & 0.47 \\
J022300.8-043408 & 0.062 & 20.11 & 0.01 & 22.97 & 0.09 & 23.59 & 0.07 & 0.69 & 0.02 & 1.98 & 0.07 & 3.05 & 0.47 \\
J022332.2-044928 & 0.058 & 18.45 & 0.05 & 22.78 & 0.34 & 24.51 & 0.23 & 1.35 & 0.10 & 6.52 & 0.81 & 6.17 & 1.20 \\
J022338.7-042431 & 0.06 & 20.69 & 0.00 & 23.36 & 0.15 & 24.20 & 0.06 & 0.83 & 0.05 & 2.01 & 0.08 & 2.88 & 0.45 \\
J022350.6-045045 & 0.059 & 17.89 & 0.14 & 23.33 & 0.44 & 25.23 & 0.18 & 1.44 & 0.18 & 11.74 & 1.11 & 10.96 & 1.94 \\
J022357.0-051918 & 0.055 & 20.20 & 0.02 & 22.93 & 0.40 & 25.14 & 0.06 & 1.61 & 0.17 & 3.87 & 0.30 & 6.18 & 1.04 \\
J022358.3-050756 & 0.059 & 20.76 & 0.01 & 24.21 & 0.06 & 24.79 & 0.03 & 0.66 & 0.02 & 2.55 & 0.05 & 2.82 & 0.43 \\
J022400.9-044943 & 0.06 & 20.00 & 0.00 & 23.51 & 0.08 & 24.39 & 0.04 & 0.85 & 0.03 & 3.02 & 0.07 & 2.82 & 0.43 \\
J022429.6-044037 & 0.057 & 18.87 & 0.17 & 24.25 & 0.80 & 25.16 & 0.75 & 0.87 & 0.07 & 7.22 & 2.32 & 6.74 & 2.39 \\
J022443.4-045158 & 0.065 & 20.01 & 0.02 & 23.33 & 0.15 & 24.57 & 0.07 & 1.07 & 0.05 & 3.26 & 0.14 & 2.77 & 0.43 \\
J022522.7-045519 & 0.064 & 18.66 & 0.01 & 22.95 & 0.14 & 24.48 & 0.09 & 1.23 & 0.04 & 5.82 & 0.27 & 4.44 & 0.70 \\
\hline
\end{tabular}
\end{minipage}

\begin{minipage}{\textwidth}
\captionof{table}{Surface brightness photometry values in the $i$ band.}
\label{tab:i_phot}
\centering
\footnotesize
\begin{tabular}{llllllllllllll}
\hline \hline
Name & $A_i$ & $m_i$ & e($m_{i}$) & $\mu_{0,i}$ & e($\mu_{0,i}$) & $\overline{\mu}_{eff,i}$ & e($\overline{\mu}_{eff,i}$) & $n_i$ & e($n_{i}$) & $r_{\text{eff},i}$ & e($r_{\text{eff},i}$) & $R_{\text{eff},i}$ & e($R_{\text{eff},i}$) \\
MGTH\_ & mag & \multicolumn{2}{c}{mag} & 
\multicolumn{2}{c}{mag arcsec$^{-2}$} & 
\multicolumn{2}{c}{mag arcsec$^{-2}$} & 
 & & \multicolumn{2}{c}{arcsec} & 
 \multicolumn{2}{c}{kpc} \\ \hline
J021534.4-042904 & 0.035 & 21.21 & 0.00 & 23.63 & 0.14 & 24.40 & 0.03 & 0.79 & 0.06 & 1.73 & 0.05 & 3.01 & 0.46 \\
J021625.5-044228 & 0.034 & 20.26 & 0.00 & 23.35 & 0.13 & 23.93 & 0.04 & 0.66 & 0.05 & 2.16 & 0.07 & 1.94 & 0.30 \\
J021642.0-044221 & 0.033 & 18.83 & 0.03 & 22.30 & 0.34 & 24.87 & 0.06 & 1.80 & 0.14 & 6.44 & 0.42 & 5.81 & 0.95 \\
J021645.4-042035 & 0.035 & 19.39 & 0.04 & 23.50 & 0.18 & 24.81 & 0.10 & 1.11 & 0.06 & 4.85 & 0.21 & 8.60 & 1.34 \\
J021714.5-052922 & 0.039 & 17.35 & 0.01 & 22.21 & 0.13 & 23.50 & 0.04 & 1.10 & 0.05 & 6.77 & 0.16 & 2.31 & 0.35 \\
J021724.3-043322 & 0.034 & 20.59 & 0.00 & 23.86 & 0.08 & 24.71 & 0.06 & 0.84 & 0.02 & 2.66 & 0.08 & 1.88 & 0.29 \\
J021759.3-041634 & 0.036 & 17.07 & 0.04 & 22.29 & 0.33 & 24.11 & 0.11 & 1.40 & 0.13 & 10.21 & 0.76 & 7.07 & 1.18 \\
J021808.2-045217 & 0.034 & 21.30 & 0.00 & 24.63 & 0.10 & 25.38 & 0.01 & 0.77 & 0.04 & 2.61 & 0.07 & 1.36 & 0.21 \\
J021818.0-043234 & 0.034 & 20.15 & 0.01 & 22.49 & 0.32 & 24.59 & 0.05 & 1.54 & 0.14 & 3.08 & 0.15 & 2.79 & 0.44 \\
J021848.1-045028 & 0.034 & 19.10 & 0.01 & 23.21 & 0.15 & 24.25 & 0.11 & 0.95 & 0.04 & 4.27 & 0.23 & 7.58 & 1.21 \\
J021934.4-042846 & 0.037 & 21.24 & 0.00 & 21.67 & 0.30 & 22.94 & 0.01 & 1.09 & 0.12 & 0.87 & 0.04 & 1.27 & 0.20 \\
J022011.5-051047 & 0.036 & 21.38 & 0.00 & 24.31 & 0.15 & 24.90 & 0.01 & 0.67 & 0.06 & 2.02 & 0.07 & 2.22 & 0.34 \\
J022026.3-045912 & 0.037 & 19.96 & 0.00 & 23.09 & 0.23 & 24.48 & 0.05 & 1.15 & 0.09 & 3.20 & 0.18 & 2.88 & 0.46 \\
J022029.4-041207 & 0.034 & 18.10 & 0.02 & 22.86 & 0.12 & 24.19 & 0.04 & 1.12 & 0.04 & 6.60 & 0.21 & 2.72 & 0.42 \\
J022042.1-052115 & 0.039 & 18.40 & 0.02 & 22.76 & 0.18 & 24.01 & 0.04 & 1.08 & 0.08 & 5.28 & 0.19 & 0.78 & 0.12 \\
J022045.0-050058 & 0.039 & 20.76 & 0.00 & 22.78 & 0.43 & 24.40 & 0.01 & 1.29 & 0.18 & 2.13 & 0.15 & 3.73 & 0.62 \\
J022051.9-045832 & 0.04 & 19.57 & 0.01 & 23.91 & 0.11 & 24.66 & 0.02 & 0.78 & 0.04 & 4.16 & 0.09 & 3.87 & 0.59 \\
J022056.9-045917 & 0.04 & 21.34 & 0.00 & 23.12 & 0.19 & 24.26 & 0.03 & 1.01 & 0.08 & 1.54 & 0.05 & 2.44 & 0.37 \\
J022058.7-051320 & 0.038 & 20.97 & 0.01 & 22.99 & 0.25 & 23.80 & 0.02 & 0.81 & 0.11 & 1.47 & 0.06 & 1.98 & 0.31 \\
J022110.7-050630 & 0.036 & 21.02 & 0.01 & 24.06 & 0.15 & 24.83 & 0.04 & 0.79 & 0.06 & 2.30 & 0.10 & 1.97 & 0.31 \\
J022117.8-045040 & 0.037 & 20.09 & 0.00 & 22.78 & 0.18 & 23.95 & 0.03 & 1.03 & 0.07 & 2.36 & 0.07 & 2.17 & 0.33 \\
J022138.1-052631 & 0.042 & 16.73 & 0.01 & 23.47 & 0.11 & 23.97 & 0.09 & 0.61 & 0.02 & 11.17 & 0.48 & 1.55 & 0.24 \\
J022145.5-050210 & 0.037 & 20.42 & 0.01 & 22.71 & 0.11 & 23.74 & 0.06 & 0.95 & 0.04 & 1.84 & 0.06 & 2.84 & 0.44 \\
J022236.8-043432 & 0.042 & 19.90 & 0.00 & 22.90 & 0.09 & 23.77 & 0.07 & 0.85 & 0.02 & 2.37 & 0.08 & 3.44 & 0.53 \\
J022300.8-043408 & 0.046 & 19.75 & 0.08 & 22.47 & 0.55 & 23.78 & 0.14 & 1.11 & 0.23 & 2.56 & 0.34 & 3.93 & 0.79 \\
J022332.2-044928 & 0.043 & 18.25 & 0.06 & 22.68 & 0.29 & 24.28 & 0.22 & 1.27 & 0.07 & 6.41 & 0.78 & 6.07 & 1.17 \\
J022338.7-042431 & 0.045 & 20.41 & 0.00 & 23.39 & 0.07 & 24.04 & 0.05 & 0.71 & 0.02 & 2.12 & 0.06 & 3.04 & 0.46 \\
J022350.6-045045 & 0.044 & 17.79 & 0.18 & 23.22 & 0.48 & 25.01 & 0.20 & 1.38 & 0.20 & 11.09 & 1.34 & 10.36 & 1.99 \\
J022357.0-051918 & 0.041 & 19.92 & 0.03 & 22.77 & 0.27 & 24.88 & 0.05 & 1.55 & 0.11 & 3.92 & 0.23 & 6.25 & 1.01 \\
J022358.3-050756 & 0.043 & 20.71 & 0.01 & 24.26 & 0.08 & 24.75 & 0.03 & 0.60 & 0.03 & 2.56 & 0.06 & 2.83 & 0.43 \\
J022400.9-044943 & 0.045 & 19.78 & 0.01 & 23.44 & 0.08 & 24.27 & 0.04 & 0.83 & 0.03 & 3.16 & 0.08 & 2.95 & 0.45 \\
J022429.6-044037 & 0.042 & 18.65 & 0.18 & 24.08 & 0.83 & 25.03 & 0.77 & 0.90 & 0.09 & 7.51 & 2.44 & 7.01 & 2.51 \\
J022443.4-045158 & 0.048 & 19.90 & 0.08 & 23.41 & 0.43 & 24.41 & 0.15 & 0.93 & 0.17 & 3.19 & 0.35 & 2.71 & 0.51 \\
J022522.7-045519 & 0.048 & 18.53 & 0.00 & 22.86 & 0.13 & 24.34 & 0.09 & 1.20 & 0.04 & 5.78 & 0.25 & 4.41 & 0.69 \\
\hline
\end{tabular}
\end{minipage}

\clearpage

\section{HI spectra of optically identified LSBGs}\label{app:hispec}

\bigskip
\centering
\includegraphics[width=\linewidth, height=0.9\textheight, 
        keepaspectratio,clip, trim={1cm 6cm 1cm 6cm}]{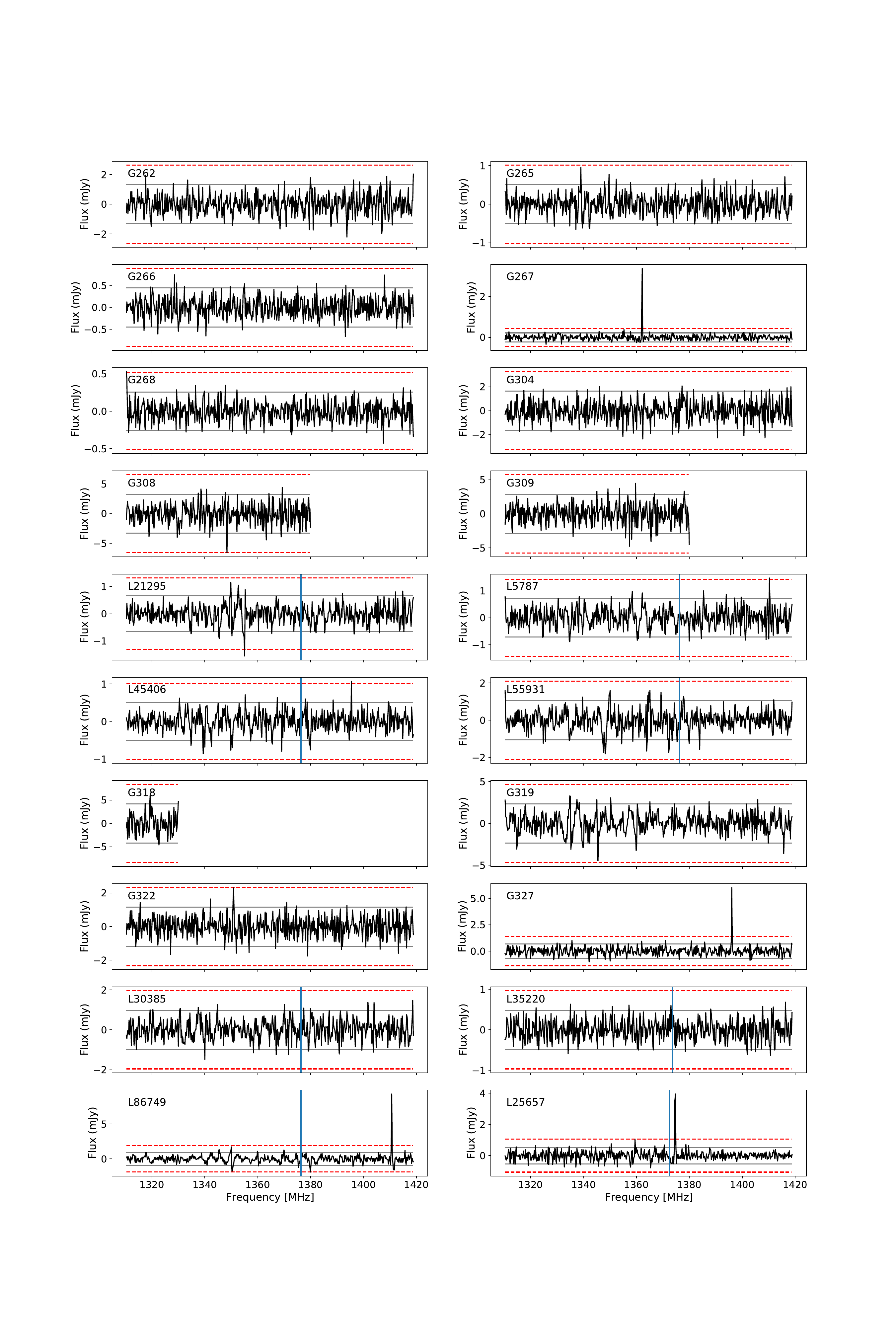}
\captionof{figure}{\hi\ spectra for every optically identified galaxy. Solid gray horizontal lines indicate $\pm$2 times the standard deviation in the spectra, and the dashed red lines indicate $\pm$4 times the standard deviation. For the L23 galaxies, the optical redshift of the assigned host galaxy is indicated by the blue lines. The G18 galaxies have no redshift information available.}
\label{fig:hi_spec}

\end{document}